\documentclass[usegraphicx,usenatbib]{mn2e}
\usepackage[a4paper,centering, totalwidth=520pt, totalheight=700pt]{geometry}
\usepackage{aas_macros}
\usepackage{amssymb}
\usepackage{amsmath}
\usepackage{microtype}
\usepackage{color}
\usepackage{float}
\usepackage{flushend}
\newcommand{\msol}{\ensuremath{\, {\rm M}_\odot}}         
\newcommand{\kpc}{\ensuremath{\, {\rm kpc}}}         
\newcommand{\mpc}{\ensuremath{\, {\rm Mpc}}}         
\newcommand{\gyr}{\ensuremath{\, {\rm Gyr}}}%

\title[Rhapsody-G I: massive galaxy clusters]
{Rhapsody-G simulations I: the cool cores, hot gas\\ and stellar content of massive galaxy clusters}

\author[O. Hahn et al.]{
Oliver Hahn\thanks{Email: oliver.hahn@oca.eu}$^{1,2}$, 
Davide Martizzi$^{3}$, 
Hao-Yi Wu$^{4,5}$,
August E. Evrard$^{5}$, \newauthor
\,\,Romain Teyssier$^{6}$ 
and Risa H. Wechsler$^{7,8}$ \\
\\
$^{1}$ Laboratoire Lagrange, Universit\'e C\^ote d'Azur, Observatoire de la C\^ote d'Azur, CNRS,\\
\quad Blvd de l'Observatoire, CS 34229, 06304 Nice cedex 4, France\\
$^{2}$ Department of Physics, ETH Zurich, CH-8093 Z\"urich, Switzerland\\
$^{3}$ Department of Astronomy, University of California, Berkeley, CA 94720-3411, USA\\
$^{4}$ California Institute of Technology, MC 367-17, Pasadena, CA 91125, USA\\
$^{5}$ Department of Physics, University of Michigan, Ann Arbor, MI 48109, USA\\  
$^{6}$ Institute for Computational Science, University of Zurich, CH-8057 Z\"urich, Switzerland\\
$^{7}$ Kavli Institute for Particle Astrophysics and Cosmology \& Physics Department, Stanford University, Stanford, CA 94305, USA\\
$^{8}$ SLAC National Accelerator Laboratory, Menlo Park, CA 94025, USA\
}
\date{draft}
\begin{document}
\maketitle


\begin{abstract} We present the {\sc Rhapsody-G} suite of cosmological hydrodynamic AMR zoom simulations of ten massive galaxy clusters at the $M_{\rm vir}\sim10^{15}\,{\rm M}_\odot$ scale. These simulations include cooling and sub-resolution models for star formation and stellar and supermassive black hole feedback. The sample is selected to capture the whole gamut of assembly histories that produce clusters of similar final mass.  We present an overview of the successes and shortcomings of such simulations in reproducing both the stellar properties of galaxies as well as properties of the hot plasma in clusters.  

\noindent In our simulations, a long-lived cool-core/non-cool core dichotomy arises naturally, and the emergence of non-cool cores is related to low angular momentum major mergers.  Nevertheless, the cool-core clusters exhibit a low central entropy compared to observations, which cannot be alleviated by thermal AGN feedback. For cluster scaling relations we find that the simulations match well the $M_{500}-Y_{500}$ scaling of {\em Planck} SZ clusters but deviate somewhat from the observed X-ray luminosity and temperature scaling relations in the sense of being slightly too bright and too cool at fixed mass, respectively. Stars are produced at an efficiency consistent with abundance matching constraints and central galaxies have star formation rates consistent with recent observations. While our simulations thus match various key properties remarkably well, we conclude that the shortcomings strongly suggest an important role for non-thermal processes (through feedback or otherwise) or thermal conduction in shaping the intra-cluster medium.
\end{abstract}



\section{Introduction}
\label{sec:intro}
While simulations of galaxy formation in Milky Way-sized haloes \citep[e.g.][]{Guedes:2011, Christensen2014} or small cosmic volumes \citep[e.g.][]{Vogelsberger:2014, Dubois:2014, Schaye:2015} are making substantial progress, realizing the population of galaxies that reside in the most massive cosmic haloes, those hosting rich clusters of galaxies, remains a formidable challenge \citep{KravtsovBorgani:2012}.  Compared to a $10^{12} \msol$ Milky Way halo, simulating a $10^{15} \msol$ cluster requires at least an extra order of magnitude in spatial resolution and three orders of magnitude in mass, thus requiring larger and longer computations. In addition, the Gaussian random field nature of the initial conditions biases the progenitors of clusters toward extreme systems at all redshifts, meaning cluster evolution is tied to that of the first stars, the earliest proto-galaxies, and the most massive quasars and their supermassive black hole interiors at high redshifts.  Finally, the deep nature of the gravitational potential well retains most of the cosmic baryonic content associated with the dark matter, implying that a complex mix of coupled hydrodynamic, magnetohydrodynamic, chemical, and radiative processes must be solved in order to gain insights into the cycling and transport of mass, energy/entropy, momentum, and metals across $\sim 10 \mpc$ regions over most of the $13.8 \gyr$ history of the universe.  

Early hydrodynamic simulations with gas cooling successfully formed multiple galaxies within a cluster \citep{KatzWhite:1993, Evrard1994}, but the central galaxies were too massive compared to observations.  Radio and X-ray observations of cavities near the central galaxy highlighted the need for strong feedback to curtail central cooling \citep[c.f.,][]{McNamara:2007, McNamara:2012}.  Inclusion of supermassive black hole (SMBH) feedback into semi-analytic methods applied to halo ensembles from N-body simulations led to a reduction in central galaxy stellar masses \citep{Croton2006, deLucia2007}.  Application within hydrodynamic simulations required estimating accretion rates onto the SMBH of active galactic nuclei (AGN), a task that in principle requires very high spatial resolution.  Schemes such as that developed by \citet{Booth:2009} were promoted to estimate Bondi--Hoyle accretion rates from simulations with roughly kiloparsec spatial resolution. Gas dynamic simulations using this approach to AGN feedback show improvements to central cluster galaxy morphology \citep{Martizzi:2012} as well as improved scaling of hot intracluster medium (ICM) properties with halo mass \citep[e.g.][and references therein]{Planelles:2014, LeBrun:2014}.  

The detailed nature of AGN feedback and its relationship to the phenomenology of the core plasma in clusters remain subjects of active investigation.  X-ray observations show that the massive cluster population can be divided into cool-core (CC) and non-cool core (NCC) categories, with the distinction determined by the strength of a surface brightness cusp within the inner $\sim 100 \kpc$ \citep{Allen:2011}.  The physics that controls this CC/NCC dichotomy has remained under debate.  Major mergers may be capable of driving a CC to NCC transition through shock heating and/or ablation of core gas, but early cosmological gas dynamic simulations with cooling and supernova feedback found major mergers to be ineffective at heating the core gas \citep{Burns:2008} unless the merger had sufficiently low angular momentum \citep{Poole:2008}.  The observational finding that morphologically disturbed clusters rarely host cool cores offers hoewever strong empirical evidence that mergers play a role in determining core morphology \citep{Pratt2010}.  
 
Still, it remains possible that internal processes, such as AGN jet-driven turbulent heating, may drive the transition from a CC to a NCC state, at least temporarily \citep{Guo:2009, Parrish:2009, Guo:2010, Parrish:2010, Ruszkowski:2010} or may play an important role in pressurising the cores.  Idealized, high-resolution AMR simulations find cold, chaotic accretion onto the SMBH at rates many times the Bondi expectation \citep{Gaspari2013}, and this has motivated models in which precipitation of turbulent core gas serves to self-regulate AGN accretion and feedback \citep{Gaspari2013b, Li:2014, Li:2015}.  \citet{Voit2015} provide empirical support that such a model, coupled with moderate conductive heating from the external reservoir of ICM plasma, may help explain CC/NCC dichotomy.  Ultraviolet imaging of brightest cluster galaxies in the CLASH cluster sample \citep{Donahue2015}  reveals knots and extended filamentary structures suggestive of the bi-polar streams that emerge in the simulations of \citet{Li:2015}. 

In this paper, we present results from a suite of gas dynamic simulations of ten high mass haloes. These simulations extend our previous suite of $N$-body simulations of massive clusters falling inside a very narrow mass range at $z=0$ \citep{Wu:2013b,Wu:2013} to multi-physics adaptive mesh simulations. They include cooling, star formation and supermassive black holes as well as their respective feedback. These subgrid models have been shown before to reproduce realistic BCG masses \citep{Martizzi:2012}. Here, we investigate how well our simulations reproduce the observed X-ray properties of galaxy clusters, most notably their density, temperature and entropy profiles and find that, much like in observed systems, our simulations produce a clear CC/NCC dichotomy most clearly seen in the entropy profiles for which we are able to give a physical explanation in terms of major cluster mergers. Reproducing the properties of the hot cluster plasma however has to go hand-in-hand with achieving also compatible results for the full galaxy population in and around clusters. While the field is still far from predictive simulations across the full multi-wavelength range covered by current and future observational studies of clusters, we are able to highlight several successes as well as important shortcomings of such state-of-the-art cluster simulations in a full cosmological context.

The structure of this article is as follows. In Section~\ref{sec:sim}, we introduce the sample of zoom simulations that we study as well as the numerical methods and models that we employ. We then compare the ICM profiles of the simulated clusters with various X-ray observations in Section~\ref{sec:icm_properties}. In Section~\ref{sec:origin_NCC}, we investigate the origin of the cool-core / non-cool-core dichotomy that we find in our sample.  We then compare the stellar properties of the galaxies formed in the simulations with observational data in Section~\ref{sec:stellar_properties}. Finally, in Section~\ref{sec:scalingrelations}, we study the evolution of the simulated clusters along several mass-observable scaling relations important for cosmology. We discuss the impact of specific modelling choices and the influence of numerical effects on our results in Section~\ref{sec:discussion}, before we summarise our results and conclude in Section~\ref{sec:conclusions}.

\section{Description of the Simulations and the numerical approach}
\label{sec:sim}
In this section, we describe how we selected a representative sample of massive clusters at fixed mass at $z=0$ from a simulation of a large cosmological volume, combining an average and a tail-biased set of objects selected from a larger sample.  We also discuss the details of our numerical and algorithmic approaches, in particular the sub-grid models we have employed to account for sub-resolution physics due to cooling and energy injection by stars and massive black holes.

\subsection{Initial conditions and general approach}
The current {\sc Rhapsody-G} simulation suite includes ten hydrodynamical zoom-in simulations selected from the original {\sc Rhapsody} sample of massive galaxy clusters \citep{Wu:2013}.  Nine are chosen to have a similar final mass of $M_{\rm vir} \approx 6 \times10^{14}h^{-1}M_\odot$ and the tenth has $M_{\rm vir} \approx 1.3 \times10^{15}h^{-1}M_\odot$. The original {\sc Rhapsody} clusters were identified from one of the {\sc Carmen} simulations from the LArge Suite of DArk MAtter Simulations ({\sc LasDamas} \footnote{http://lss.phy.vanderbilt.edu/lasdamas/}). The {\sc Carmen} simulation on which we base our re-simulations is a cosmological volume of $1\,h^{-1}{\rm Gpc}$. All simulations are based on a $\Lambda$CDM cosmology with density parameters $\Omega_b=0.045$ for baryons and $\Omega_m=0.25$ for total matter, $\Omega_\Lambda=0.75$ for the cosmological constant. The long-wave spectral index is $n_s=1$, the amplitude normalization is $\sigma_8=0.8$, and the Hubble parameter is $h=0.7$. The original {\sc Rhapsody} haloes were identified based on a {\sc Gadget-2} \citep{Springel2005b} simulation of the full box using $1120^3$ particles. The subsequent $N$-body only {\sc Rhapsody} simulations were also carried out using {\sc Gadget-2} but using ``zoom'' initial conditions \citep[using {\sc MUSIC};][]{Hahn2011a} for a sphere of $8\,h^{-1}{\rm Mpc}$ centred on each selected cluster at $z=0$ with an effective resolution of $4096^3$ (4K) and $8192^3$ (8K) particles \citep[see][ for details]{Wu:2013}. All initial conditions (both for the original box and all subsequent zooms) were performed using second-order Lagrangian perturbation theory at $z=50$. 

In this paper, we re-generate the respective initial conditions using {\sc Music} for ten clusters (see Section~\ref{sec:RGsubset} below for details on the selection and properties of these haloes) but now including baryon perturbations. We assume here that baryons are fully tracing the dark matter perturbations already at $z=50$, which is accurate enough for our purposes here (and common procedure), but strictly speaking not correct in detail \citep[see e.g.][ for a detailed discussion]{Angulo2013}. In addition, we use also Lagrangian perturbation theory for the baryons using the local Lagrangian approximation \citep[see][ for details on baryon ICs for grid and particle codes]{Hahn2011a}.

\subsection{The {Rhapsody-G} subset of {Rhapsody} clusters}
\label{sec:RGsubset}
From the sample of $\sim$100 clusters in the $M_{\rm vir}=10^{14.8\pm0.05}\,h^{-1}{\rm M}_\odot$ mass bin of the original {\sc Rhapsody} sample \citep{Wu:2013}, we selected a subset of 10 clusters which we investigate using multi-physics simulations in this paper. To sample both extreme cases of formation history and more average clusters, we considered the two-dimensional ordering shown in Figure~2 of \cite{Wu:2013}, where clusters are ranked first by halo concentration $c$, and then at similar concentration ranked a second time by the numbers of subhaloes $N_{\rm sub}$ above $v_{\rm max}>100\,{\rm km/s}$. The extreme corners of this space are occupied by the systems with IDs 337, 377, 572 and 653. We additionally included system 545 which has similar properties to 337 (high concentration, high substructure fraction). We note that this naturally includes the fossil system 572 which was discussed in more detail already in \cite{Wu:2013b} as being a particularly early forming system and occupying the tail in many halo properties. We complement this subset of ``extreme'' clusters with four more clusters taken from the central region of the $c$-$N_{\rm sub}$ space. These are the ones with IDs 211, 348, 361, 448. Additionally we include cluster 474, which has a mass that is twice as high as the rest of the sample. These clusters, along with various properties discussed in Section~\ref{sec:global_properties} are listed in Table~\ref{tab:rhapsody_g_sample}.

\subsection{Hydrodynamics, N-body and gravity}
In order to follow the non-linear multi-physics evolution of our initial conditions, we use here the Eulerian adaptive mesh refinement code {\sc Ramses} \citep{Teyssier:2002} and include radiative cooling, as well as sub-grid models for star formation and AGN feedback which we discuss below. {\sc Ramses} is based on a MUSCL scheme with an approximate Riemann solver and gravity is solved using the multi-grid method directly on the AMR hierarchy. Dark matter as well as stellar and sink particles are evolved using standard $N$-body techniques.

We employ an overdensity-based (i.e. ``Lagrangian'') refinement strategy that splits cells if they reach an overdensity of eight, i.e. refinement of the base grid by $n$ additional levels requires a density of $8^n\bar{\rho}$. The maximum refinement level for the 4K runs (which constitute the main part of our analysis here) is determined by maintaining a smallest cell size of physical $3.8\,h^{-1}{\rm kpc}$. The 4K dark matter $N$-body particle mass is $8.27\times10^8\,{h^{-1}{\rm M}_\odot}$, and the initial mass per hydro cell is $\sim1.82\times10^8\,{h^{-1}{\rm M}_\odot}$ respectively. The 8K run of RG~653, which we consider here for convergence purposes, has twice the mass and force resolution, i.e. $1.9\,h^{-1}{\rm kpc}$, and an eight times smaller $N$-body particle mass than the 4K runs. The high-resolution Lagrangian patch from which the $8\,h^{-1}{\rm Mpc}$ sphere centred on each cluster will form is tagged using a passive scalar colour field that is advected with the gas. Dynamic refinement is restricted to the regions where this colour field is non-zero. We thus focus all computational resources on the forming cluster and its immediate environment.

\subsection{Modelling cooling, star formation, stellar feedback and chemical evolution}
The simulations include optically thin radiative cooling using the cooling rates of \cite{Sutherland:1993} for H, He and metal line cooling. Different metals are not evolved separately, but the total gas metallicity is advected with the hydrodynamical equations as a passive scalar and is sourced by the supernova feedback model. A UV background is included assuming the parameterisation of \cite{Haardt:1996} and an instantaneous reionisation at $z=10$ taking into account an earlier reionisation in the particularly overdense proto-cluster environment.

The unresolved cold and dense gas that will constitute the interstellar medium (ISM) of galaxies is approximated using a polytropic temperature floor
\begin{equation}
T_{\rm floor}=T_*\left( \frac{n_{\rm H}}{n_*} \right)^{\gamma_* -1}
\end{equation}
where $n_*=0.1$ cm$^{-3}$ is the threshold for star formation (see below), $T_*=10^4$ K is a characteristic temperature mimicking thermal and turbulent motions in the ISM, and $\gamma_*=5/3$ is the effective polytropic exponent. In practice, gas can be heated above the temperature floor, but cannot cool below it.

We assume star formation to occur in a cell when the gas density exceeds $n_*$. In this case a star particle is spawned carrying 20 per cent of the mean baryon mass of the cell. We set the local star formation rate per cell as
\begin{equation}
\dot{\rho}_*=\epsilon_*\frac{\rho_{\rm gas}}{t_{\rm ff}}
\end{equation}
with the free-fall time $t_{\rm ff}=(3\pi/32 G \rho)^{1/2}$ of the cell and a star formation efficiency per free-fall time $\epsilon_*=0.02$. Stellar particles are seeded locally from a Poisson process. 

Supernova feedback is implemented based on the model of \cite{Dubois:2008}, in which each newly formed star particle releases a fraction $\eta=0.1$ of its mass and metals with a yield of $y=0.1$ into its surrounding cells through supernovae (SNe) after $10\,{\rm Myr}$ (this implies that 1 per cent of the time integrated global star formation rate is returned as metals to the ISM). In addition, each supernova injects an energy of $10^{51}\,{\rm erg}$ into the surrounding ISM which regulates the star formation efficiency at galaxy scale halo masses. The free parameters of the supernova feedback have been calibrated to reproduce stellar masses consistent with abundance matching results \citep[e.g.][]{Behroozi2013} at halo masses $M_{\rm halo}\lesssim 10^{12}\,{\rm M}_\odot$ for haloes resolved with $\gtrsim 1000$ particles.

\subsection{Modelling AGN feedback}
\label{sec:agn_modelling}
The deep potential wells of galaxy clusters require a stronger heating source than supernova feedback to prevent a central cooling catastrophe. Active galactic nuclei (AGNs) provide a natural source of energy in massive galaxies to offset these extreme cooling flows \citep[see e.g.][for reviews]{McNamara:2007,Fabian:2012}. In the simulations discussed in this paper, we include a purely thermal AGN feedback, based on the subgrid models of \cite{Springel2005c} with the additional energy injection thresholding of \cite{Booth:2009}, which is commonly employed in SPH simulations \citep[e.g.][for recent examples]{LeBrun:2014,Schaye:2015}. We give a brief summary of our AGN feedback model below but refer the reader to Paper 2, Martizzi et al. 2015, in prep., for a more detailed exposition.

\subsubsection{Thermal AGN feedback model} 
\label{sec:agn_modelling_thermal}

Sink particles, representing supermassive black holes, are created when contiguous regions of high density gas exceed $10^{-29}\,{\rm g/cm^3}$ \citep[identified on the fly by the clustering method of][]{Bleuler:2014} that (1) do not already contain a sink particle, (2) are gravitationally bound, and (3) have an accretion rate $\dot{M}_{\rm clump}=\frac{M_4}{t_{\rm ff}} > 30 {\rm M_{\odot}/yr}$, where $M_4$ is the mass contained within a sphere of 4 computational cells. This guarantees that sink particles are spawned only in the most massive haloes at high redshift. The initial black hole mass is chosen proportional to the clump accretion rate times the Salpeter time.

Sink particles then accrete at a boosted Bondi-Hoyle accretion rate \citep[e.g.][]{Booth:2009}
\begin{equation}
\dot{M}_{\rm BH}=4\pi\alpha_{\rm boost}\tilde{r}_B^2\rho\sqrt{u^2+c_{\rm s}^2},
\end{equation}
where $u$ is the fluid velocity, $c_s$ the sound speed, $\tilde{r}_B =\min(r_B,4\Delta x)$ the free-fall limited Bondi radius, and
\begin{equation}
\alpha_{\rm boost} = \left\{ \begin{array}{cl} 
\left( \frac{n_{\rm H}}{n_*} \right)^2 & {\rm if }~n_{\rm H} > n_* = 0.1~{\rm H/cc,}\\
1 & {\rm otherwise}
\end{array}\right.
\end{equation}
the density dependent boost factor. We additionally impose the Eddington limit onto $\dot{M}_{\rm BH}$ assuming that $\epsilon_r=0.1$ of the accreted rest mass energy is converted into radiation. Numerically, the amount $\dot{M}_{\rm BH}\Delta t$ up to a maximum of half of the gas mass contained in each cell is removed from cells inside the sink radius. The thermal energy $\Delta E =  \epsilon_c \epsilon_r \dot{M}_{\rm BH} c^2 \Delta t$ would in principle be released into the gas in each time step, but following \cite{Booth:2009}, we accumulate $E_{AGN}=\sum \Delta E$ and inject the accumulated $E_{\rm AGN}$ into the sink radius (i.e. the sphere of radius 4 cells) only once $E_{\rm AGN} > \frac{3}{2} m_{\rm gas}k_B \Delta T$, where $m_{\rm gas}$ is the gas mass enclosed by the sink radius. The coupling efficiency $\epsilon_c\simeq0.15$ has been calibrated to reproduce the $M_{\rm BH}-\sigma$ relation \citep{Teyssier:2011}. In our fiducial model, we set $\Delta T = 10^7\,{\rm K}$, but vary this by one order of magnitude up and down when we investigate the effect of trading fewer violent AGN events against more frequent less violent AGN events in Section~\ref{sec:agn_impact}. We emphasise however that the total energy injected by the AGN obviously remains more or less the same. The parameter ${\Delta}T$ has been shown to allow a tuning of the bulk properties of the ICM \citep[e.g.][]{LeBrun:2014} in SPH simulations.

\subsubsection{Phenomenological AGN feedback model} 
\label{sec:agn_modelling_pheno}

Since the AGN energy is thus injected close to the resolution limit into a region containing the densest cells of the simulation, we also consider a more phenomenological model, inspired by the AGN model of \cite{Battaglia:2010}, in which we distribute $E_{\rm AGN}$ over a resolved sphere of a radius determined by the black hole mass. \cite{Battaglia:2010} used a formula for the injection radius that depends on the halo mass. Since we do not track halo masses on the fly with our algorithm and prefer a local criterion, we decided to express the injection radius used by \cite{Battaglia:2010} in terms of the black hole mass by using the relation $M_{\rm BH}\propto V_{\rm c}^4$, where $V_{\rm c}$ is the circular velocity of the halo \citep[][noting that for an NFW profile the circular velocity is related to halo mass as $V_{\rm c}\propto M_{200}^{1/3}$]{Volonteri:2011}. The radius of the injection sphere is then given by
\begin{equation}
\begin{aligned}
 R_{\rm AGN}=& {\rm max}\Big[4\Delta x(1+z), \\
  & 100 \hbox{ kpc}/h \times E(z)^{-2/3} \left(\frac{M_{\rm BH}}{7.6\times 10^{10} \hbox{M}_{\odot}}\right)^{1/4}\Big], \\
\end{aligned}
\end{equation}
where $E(z)$ is the ratio of the Hubble constant at redshift $z$ and $z=0$, i.e. $H_{\rm 0}$,
\begin{equation}
 E(z)=\frac{H(z)}{H_{\rm 0}}=\sqrt{\Omega_{\rm m}(1+z)^3+\Omega_{\Lambda}}.
\end{equation}
$R_{\rm AGN}$ is larger than 4 cell sizes at most times. This additionally reduces the thermal energy inserted into each individual cell, reducing immediate loss of energy through efficient cooling at the highest densities. Again, we only consider this alternative model, which we term the ``phenomenological model'', when investigating the impact of changes in the thermal feedback model parameters in Section~\ref{sec:agn_impact}.


\subsection{Halo/galaxy finding and their properties}
In order to identify haloes, including their stellar, black hole and gaseous content in our simulation, we use a heavily modified version of {\sc Rockstar-galaxies}, which is a special version of the original version of {\sc Rockstar} \citep{Behroozi:2013b} adapted to work also for simulations including gas and stars \citep[see e.g.][where this version was used]{Knebe:2013}. In order to interface {\sc Ramses} with {\sc Rockstar-galaxies}, we convert all leaf cells of the AMR tree to pseudo-particles of variable mass, which is however typically not varying over more than an order of magnitude due to the Lagrangian refinement strategy.

During the (sub-)halo finding, we then calculate all relevant halo and galaxy properties directly inside {\sc Rockstar-galaxies}. Unbinding of gravitationally unbound particles and cells is performed, and we obtain the masses, radii, centres, and bulk velocities of the dark matter, stellar, gaseous, and black hole content of each (sub-)halo. We use the {\sc Consistent-Trees} code \citep{Behroozi:2013c} to link the halo/galaxy catalogues across redshifts based on only the dark matter particle IDs.

We additionally compute several galaxy-related quantities, such as the star formation rate, mean stellar age, surface brightness, and magnitude in various photometric bands. For the star formation rate, we simply sum the mass of all star particles younger than 100~Myr, divided by 100~Myr. For the surface brightness and magnitude calculations, we determine the luminosity of each star particle in a given filter applied to the spectral energy distribution obtained from the simple stellar population models from {\sc Starburst99} \citep{Leitherer:1999}, assuming the metallicity and age of the star particle as the mean age and metallicity of the stellar population. In order to derive the surface brightness-limited galaxy masses, we perform a one-dimensional projection and integrate the enclosed stellar mass out to a given surface brightness limit. We note that we do {\em not} take dust absorption into account, which might affect our comparison with observations. 

Finally, we also calculate several properties related to the hot plasma, such as the X-ray luminosity and an X-ray emissivity per cell, in order to allow us to weight quantities by their observability in X-ray observations, based on an APEC emission model, similar to \cite{Biffi:2012}. We neglect however additional sophistications such as actual photon sampling, PSF effects or spectral fitting in order to determine the X-ray temperature, which are clearly beyond the scope of this paper. Finally, we also compute the SZ flux decrement for each cluster. We detail how we perform X-ray and SZ measurements separately in Sections~\ref{sec:szmasses} and \ref{sec:xrayobservables}, respectively, in the context of the evolution of the clusters along scaling relations.


\begin{figure*}
\begin{center}
\includegraphics[width=1.8\columnwidth]{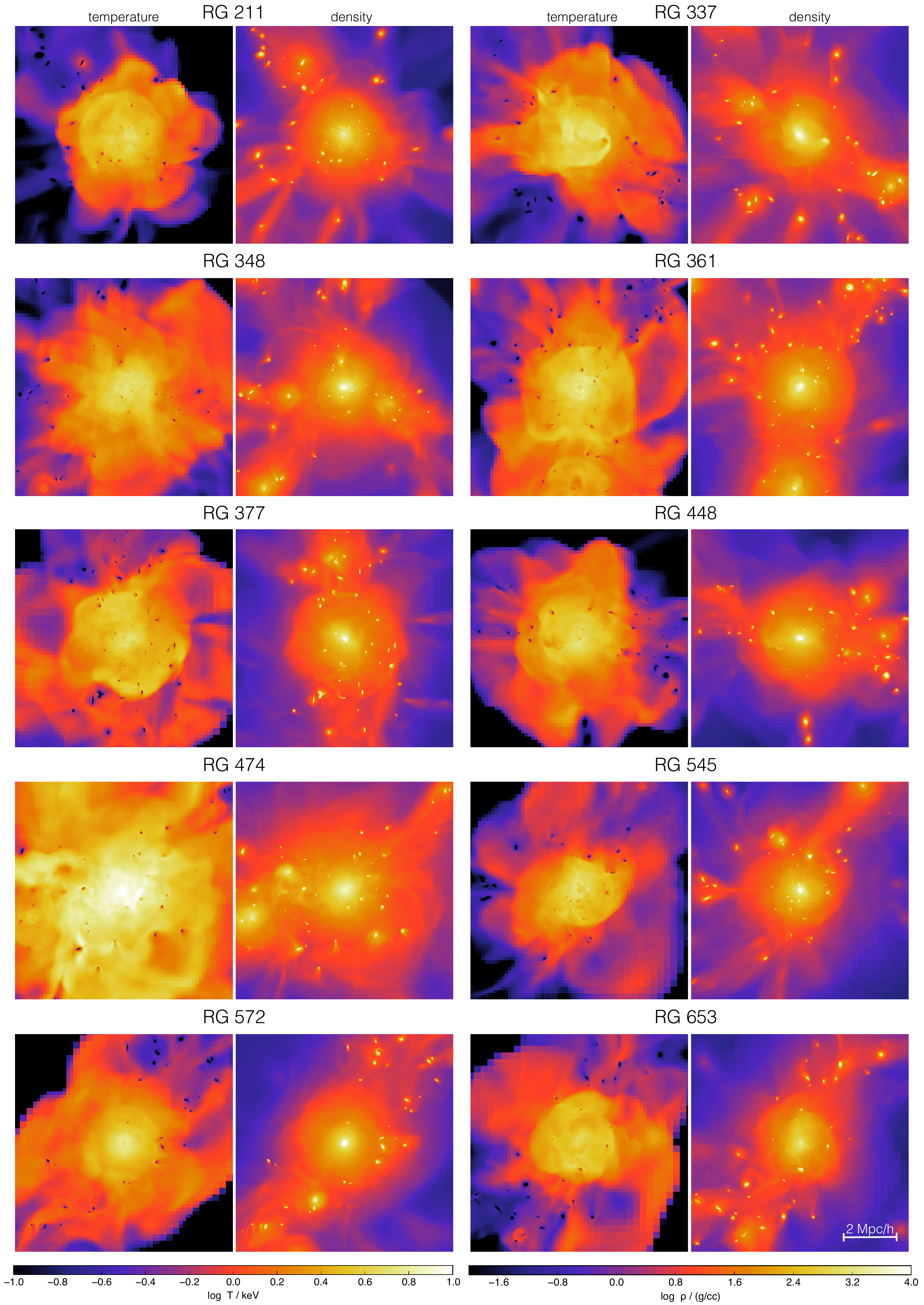}
\end{center}
\caption{\label{fig:RG_sample}Mass weighted temperature and gas density maps at $z=0$ of the simulated clusters used in this paper at the 4K resolution, i.e. $\left<\rho\right>=\int\rho^2\,{\rm d}l / \int\rho\,{\rm d}l$ and  $\left<T\right>=\int T\rho\,{\rm d}l / \int\rho\,{\rm d}l$. Notable extreme cases are: RG~337 is a merging system and RG~572 is a fossil cluster, RG~474 has twice the mass of the other systems; RG~337 and RG~377 are richer in substructure than average and RG~653 is poorer than average. The images are $8\,h^{-1}{\rm Mpc}$ wide and the projection depth is $4\,h^{-1}{\rm Mpc}$ centred on each cluster.}
\end{figure*}


\section{Comparison of ICM profiles with observations}
\label{sec:icm_properties}
We perform a detailed comparison of density, temperature, entropy, and gas depletion profiles with X-ray observations in this section. To show both the mean and deviations over cosmic time of the simulated profiles, we consider individually stacked profiles over time for each simulated cluster for all snapshots for which the cluster mass falls into a narrow range. We discuss this stacking procedure in detail below.


\subsection{Global cluster properties: masses and cool cores}
\label{sec:global_properties}
We first discuss some global properties of our sample of simulated galaxy clusters. To give a visual impression of the diversity of the sample, Figure~\ref{fig:RG_sample} shows density-weighted temperature and density maps for all clusters at $z=0$. In Table~\ref{tab:rhapsody_g_sample}, we list all the simulated clusters including their $z=0$ masses and their half-mass redshift $z_{1/2}$, i.e. the highest redshift at which the most massive progenitor exceeds half the final mass. We quote two $z=0$ mass measurements: the virial mass (calculated using the overdensity criterion of \citealt{Bryan:1998}) and $M_{500c}$, the mass enclosed inside 500 times the critical density. 

\begin{table}
\begin{center}
\begin{tabular}{lcccccc}
\hline
cluster  & $M_{vir,0}$ & $M_{500c,0}$ & $z_{1/2}$ & $N_{\rm stack}$ & $\bar{z}_{\rm stack}$ & CC? \\
\hline
RG 211 & $1.00$ & $5.02$ & 0.27 & 7 & 0.07 & -\\
RG 337 & $1.06$ & $6.59$ & 0.71 & 20 & 0.37 &-\\
RG 348 & $1.15$ & $6.28$ & 0.67 & 22 & 0.30 & -\\
RG 361 & $1.07$ & $5.46$ & 0.65 & 19 & 0.25 & + \\
RG 377 & $1.08$ & $4.89$ & 0.45 & 7 & 0.10 & + \\
RG 448 & $1.03$ & $5.19$ & 0.73 & 12 & 0.18 &+\\
RG 474 & $2.69$ & $10.38$ & 0.47 & 4 & 0.61 &-\\
RG 545 & $0.93$ & $5.12$ & 0.77 & 8 & 0.08 & + \\
RG 572 & $0.96$ & $5.65$ & 1.11 & 30 & 0.37 &(+)\\
RG 653 & $0.84$ & $3.75$ & 0.21 & 5 & 0.07 &-\\
\hline
& $\times10^{15}{\rm M}_\odot$ & $\times10^{14}{\rm M}_\odot$ & & \\
\hline
\end{tabular}
\end{center}
\caption{\label{tab:rhapsody_g_sample}The haloes selected from the original RHAPSODY sample for the RHAPSODY-G simulations listed along with the $z=0$ masses and the formation redshift $z_{1/2}$. We also list the number of snapshots that enter our stacking analysis and the mean redshift of the stack, as well as whether the cluster is classified as cool-core in the stack analysis (see Section~\ref{sec:global_properties} for details). The system RG~572 is a fossil cluster, a significant outlier in almost all its halo properties, but is close to a cool-core system. }  
\end{table}

For the comparison of the cluster ICM profiles with observational data from the ACCEPT sample \citep{Cavagnolo:2009}, in Section~\ref{sec:icm_properties}, we decided to select a very narrow mass bin $4 < M_{\rm 500c}(z) / 10^{14}\,M_\odot < 6$ in both the ACCEPT sample and the simulated clusters. For the simulated clusters, we take every fifth snapshot out of the $\sim300$ snapshots that we have stored for each cluster for which it falls into this mass bin. We then perform a stack analysis for these snapshots separately for each cluster in order to quantify the mean profiles and the variance around these mean profiles (see Section~\ref{sec:gas_xray} for details). In Table~\ref{tab:rhapsody_g_sample}, we thus also give the number of snapshots $N_{\rm stack}$ used in the stack analysis of each cluster and the mean redshift $\bar{z}_{\rm stack}$ of the stack. $N_{\rm stack}$ is most of the time, but not always, a larger number if the cluster has a high formation redshift. Finally, we indicate whether the cluster is classified as a cool-core (CC) or non-cool-core (NCC) cluster based on whether the mean central entropy of the stack entropy profile is below $40\,{\rm keV}\,{\rm cm}^2$ at $r=10\,{\rm kpc}$. We find that RG~361, RG~377, RG~448 and RG~545 are classified as cool core systems according this classification. Halo RG~572 is close to a CC in principle as well, but we will treat it separately as its core properties are even more extreme. The other systems are NCC clusters. We note that we see no immediate connection between either $z_{1/2}$ or the time the cluster spends in the $4 < M_{\rm 500c}(z) / 10^{14}\,M_\odot < 6$ mass bin and the CC/NCC distinction. Two CC clusters are from the ``extreme'' sample and two from the ``average''. The ratio of 4 out of 9 (excluding the fossil system RG~572 as an extreme outlier) is, within the errorbars of small-number statistics, consistent with the number of cool core systems in the ACCEPT sample (6 out of 28) and with other observational estimates at these mass scales \citep[e.g.][]{Chen:2007} of $\sim40$ per cent.


\begin{figure}
\begin{center}
\includegraphics[width=0.9\columnwidth]{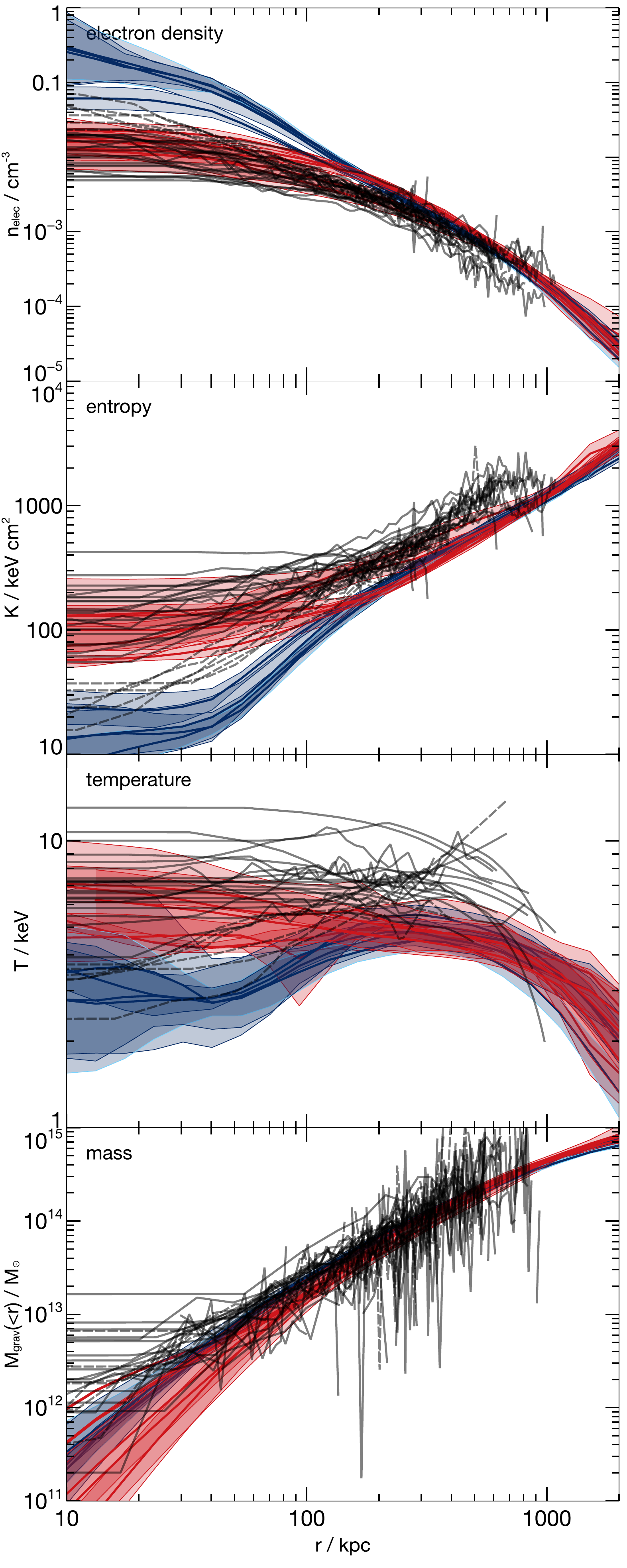}
\end{center}
\caption{\label{fig:accept_profiles}Comparison of ICM profiles from {\sc Rhapsody}-G (red and blue ribbons) with observational data from our mass-matched subset of ACCEPT clusters (black solid and dashed lines). {\em Top:} electron density profiles, {\em 2nd from top:} entropy profiles, {\em 3rd from top:} temperature profiles, and {\em bottom:} total mass profiles. The cool-core clusters in the ACCEPT subset are indicated by dashed lines, while the cool-core {\sc Rhapsody}-G clusters are shown in blue and the non-cool-core in red. The simulated haloes were stacked individually for each cluster while occupying the same mass range as the ACCEPT subset. The shaded ribbons indicate the $1\sigma$ scatter in each stack, reflecting time variations in the profiles. We excluded the fossil cluster RG~572 (see text).
}
\end{figure}

\subsection{Comparison with ICM profiles from X-ray data}
\label{sec:gas_xray}
We next compare the gas temperature, entropy, and electron density profiles to observational data. We specifically confront the ICM X-ray profile data from the ACCEPT sample \citep{Cavagnolo:2009}. As mentioned above, in order to enable a precise comparison with our {\sc Rhapsody}-G clusters, we perform a stringent mass cut of the ACCEPT clusters. We combine X-ray and SZ masses to this end to reduce the influence of hydrostatic mass bias at least to some degree. Specifically, for each cluster in the ACCEPT sample, we find its $M_{500}$ masses from the MCXC catalogue \citep{Piffaretti:2011} and its $M_{\rm SZ}$ mass from the {\em Planck} 2015 catalogue \citep{Planck:SZ:2015}. Cross-matching the three catalogues and selecting all clusters with $4\times10^{14} < M / M_\odot < 6\times10^{14}$ in both $M_{500}$ and $M_{\rm SZ}$, we finally obtain a sample of 29 ACCEPT clusters (see Table~\ref{tab:accept_sample}, all but 5 from the Abell catalogue). We additionally excluded Abell~2069 which makes the mass cuts but is a strong outlier in both its entropy and electron density profile as it is a merging cluster. As for the RG clusters, we define cool-core as having a decreasing profile and a central entropy of at most $40\,{\rm keV}\,{\rm cm}^2$ at $r=10\,{\rm kpc}$. We note that the overall trends of the ACCEPT CC profiles are in good enough agreement with the more recent analysis of \cite{Mantz:2015} of cool-core systems, most notably the slope of the entropy profiles, for our purposes here.

In Figure~\ref{fig:accept_profiles}, we show the comparison between the mass-selected ACCEPT sub-sample and stacked profiles from the {\sc Rhapsody} sample. For each simulated cluster, we show the mean profile as well as the standard deviation around that mean profile, obtaining one ribbon for each {\sc Rhapsody} cluster in Figure~\ref{fig:accept_profiles}. For this analysis, we excluded the fossil cluster RG~572 since it shows an extremely fluctuating entropy and core density, caused by a very active and rapid growing central hypermassive black hole of $\sim3\times10^{11}\,{\rm M_\odot}$ at $z=0$. Note that in the full {\sc Rhapsody} sample of more than 100 clusters this was still a completely abnormal case. This cluster is so atypical in its core properties already in pure $N$-body simulations that we only note here that it is also an outlier in its baryonic core properties. 

The number of profiles stacked for each cluster is different and given in Table~\ref{tab:rhapsody_g_sample}.  In all cases, {\em we find that the dispersion in each stack around the mean profile is smaller than the difference between the cool and non-cool-core profiles, indicating a stable bimodality}.  This can be clearly seen from the small extent of the ribbons in Figure~\ref{fig:accept_profiles}, which indicate a very small amount of scatter around the mean in the stack of each cluster over time. The cool cores are thus a much more significant feature than short term fluctuations in the profiles. The non-cool-core profiles from ACCEPT and {\sc Rhapsody} agree well within their respective scatter, for both the electron density profiles and the entropy profiles. The ACCEPT cool-core clusters however show a much weaker cool core than the {\sc Rhapsody} CC clusters: the observed clusters show only a moderate increase in core electron density inside the innermost $\sim50\,{\rm kpc}$, while the {\sc Rhapsody} cool cores show a strong drop in entropy and increase in electron density already at scales of $\lesssim150\,{\rm kpc}$. It is clear that the cool core systems are most likely still undergoing overcooling to some degree despite the central AGN that is efficiently fuelled during the CC phase. The CC/NCC dichotomy is thus a long lived property of our clusters, consistent with the observational constraints of e.g. \cite{McDonald:2013}. The dichotomy arose naturally in a larger sample of cosmological cluster simulations and can be explained by differences in the assembly history and nature of major mergers of the clusters (see our analysis in Section~\ref{sec:origin_NCC}).

Interestingly, outside the core, the temperature profiles show a somewhat discrepant temperature at large radii, specifically a $\sim 30$ per cent difference in X-ray temperature at $\gtrsim200\,{\rm kpc}$, with much better agreement at smaller radii. Upon closer inspection one notices that a similar but slightly weaker offset also exists in the power-law part of the entropy profile, where the simulated profiles are systematically offset to lower entropy. A similar discrepancy can be seen also, e.g., in Figure~7 of \cite{Dubois:2011}, independently of the much larger range of subgrid models employed in that reference. This may point towards additional physics missing from these simulations.

\begin{figure}
\begin{center}
\includegraphics[width=0.8\columnwidth]{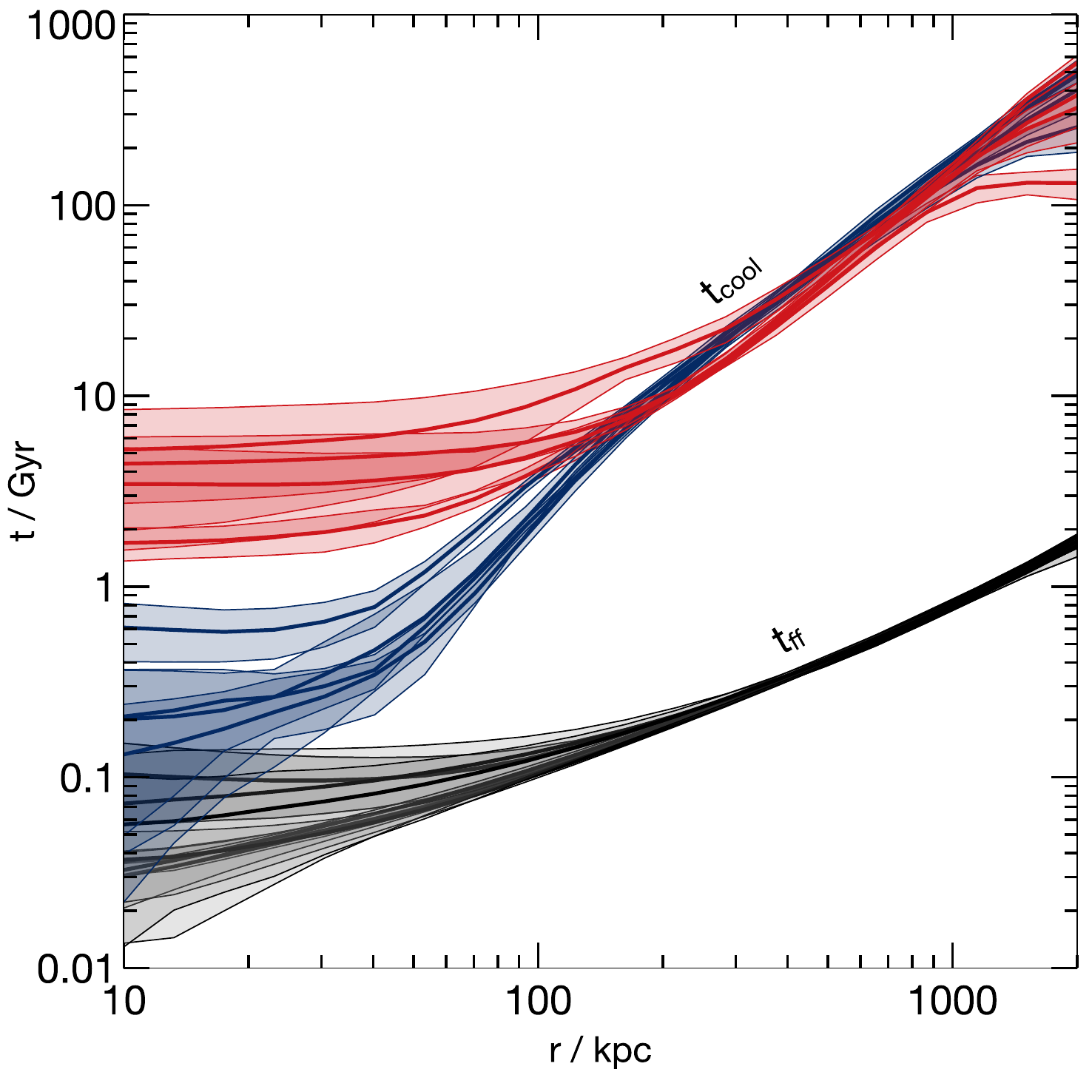}
\end{center}
\caption{\label{fig:times}Comparison of the cooling time profiles of cool-core (blue) and non-cool-core (red) clusters with the respective free-fall-time profiles (gray). The profiles have been stacked individually for each cluster as in Fig.~\ref{fig:accept_profiles}. All cool-core clusters have a cooling time below $\sim1\,{\rm Gyr}$ in their cores.}
\end{figure}

In order to investigate better the nature of the cool cores, in Figure~\ref{fig:times}, we plot cooling and free-fall time profiles \citep[c.f. also Fig. 9 of][]{Li:2015}. We see that for the non-cool core systems, the cooling time is {\em at least} a factor of 10 above the free-fall time in the core, while for the cool core systems, the two time scales are much closer to each other. Normally the central AGN should increase the central cooling time dramatically, but it appears to not prevent the formation of cool cores in four of our systems. This is despite a dramatic growth that the central black hole undergoes in the CC cases. For one of them, RG~545, we demonstrate in Section~\ref{sec:agn_impact} that the existence of the cool core does {\em not} depend on the details of the AGN feedback parameters. {\em We are thus led to believe that in our simulations the CC/NCC dichotomy has a cosmological origin, and will investigate this aspect further in Section~\ref{sec:origin_NCC}.} 


\subsection{Hydrostatic mass and deviations}
\label{sec:hydrostaticmass}

\begin{figure}
\begin{center}
\includegraphics[width=0.85\columnwidth]{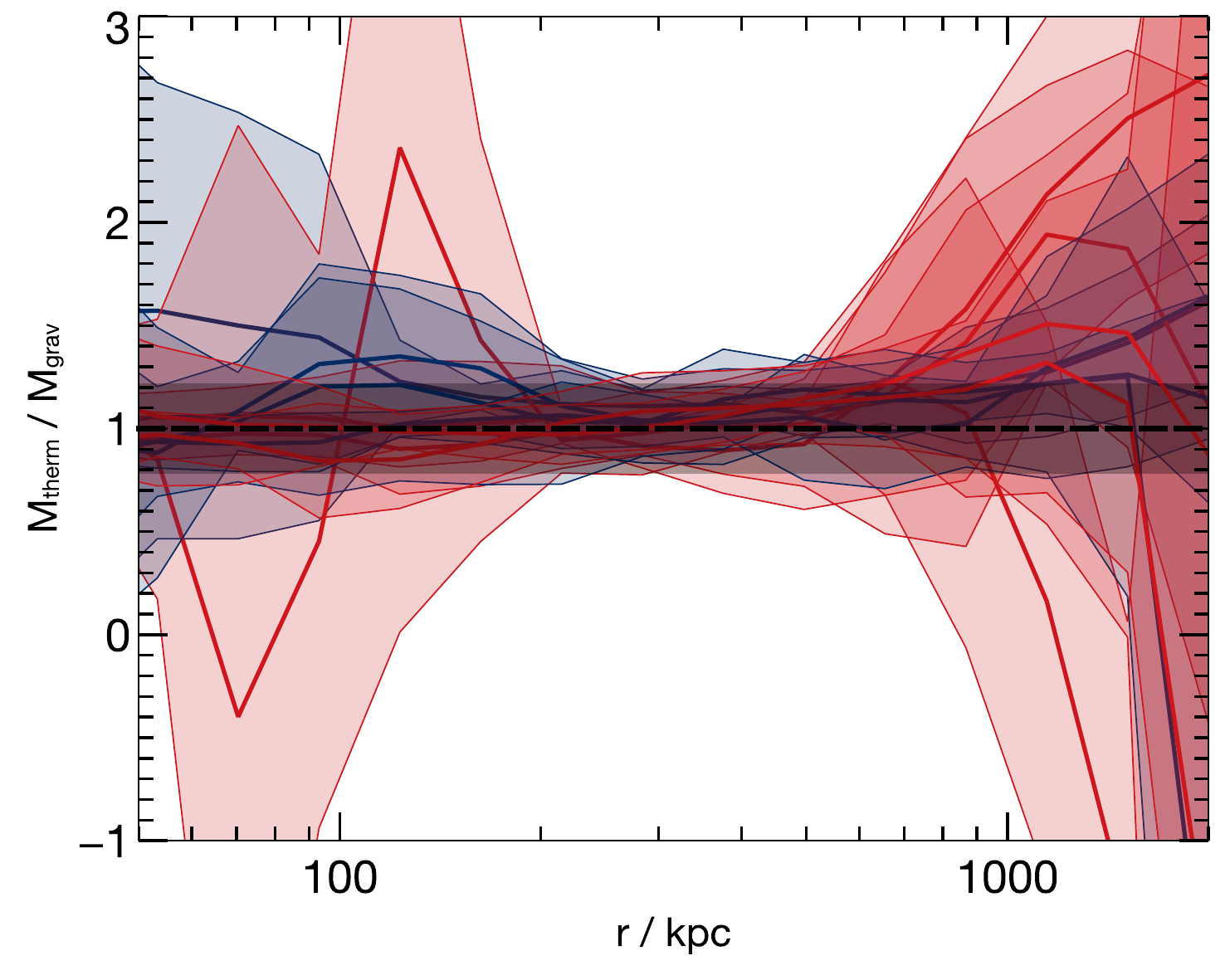}
\end{center}
\caption{\label{fig:hydrostaticmass}Ratio of the hydrostatic and the gravitating mass as a function of radius. Outside of the core and well inside the virial radius, the clusters are very close to hydrostatic equilibrium, with large deviations at radii $\gtrsim1\,{\rm Mpc}$. There is no obvious difference between the CC systems (blue) and the NCC systems (red). The grey shaded band indicates $\pm20$ per cent deviations from hydrostatic equilibrium.}
\end{figure}

The cluster is stabilised against gravitational forces by thermal pressure as well as bulk and random flow motions \citep[e.g.][]{Nelson:2014}. We next investigate to what degree the thermal pressure alone counteracts gravity in our simulations. In adaptive mesh simulations with a Lagrangian refinement strategy (which is however indispensable if galaxy formation is included), the numerical Reynolds number is relatively high so that no turbulent cascade develops and the energy is very quickly dissipated into thermal energy. In contrast to uniform high resolution simulations with fully developed turbulence \citep{Miniati:2015}, we can focus on thermal effects here inside the outer infall region. We follow \cite{Nelson:2014} and define the thermal (or hydrostatic) mass as
\begin{equation}
M_{\rm therm}(r) = \frac{-r^2}{G\left<\rho\right>}\frac{\partial\left<P\right>}{\partial \,r},
\end{equation}
where the angle brackets denote mass-weighted averages over spherical shells at a given radius $r$. In Figure~\ref{fig:hydrostaticmass}, we show the ratio $M_{\rm therm}(r)/M_{\rm grav}(r)$, where $M_{\rm grav}(r) = M_{\rm tot}(<r)$ is the total mass enclosed in radius $r$. All clusters are closest to hydrostatic equilibrium at radii of $\sim300\,{\rm kpc}$ with a mean deviation of at most 20 per cent inside of $\sim600\,{\rm kpc}$. At large radii, the variance is however substantial, most likely due to incomplete thermalisation of the region which is perturbed by accreting gas. Most importantly for our analysis, {\em we find no significant difference in deviations from hydrostatic equilibrium over radii $100\,{\rm kpc}\lesssim r \lesssim1\,{\rm Mpc}$ between CC and NCC systems}. There is however an indication of systematic differences at $\gtrsim1\,{\rm Mpc}$, as well as, obviously, at scales of the core.

\subsection{Gas depletion profiles}
\label{sec:gas_depletion}

\begin{figure}
\begin{center}
\includegraphics[width=0.9\columnwidth]{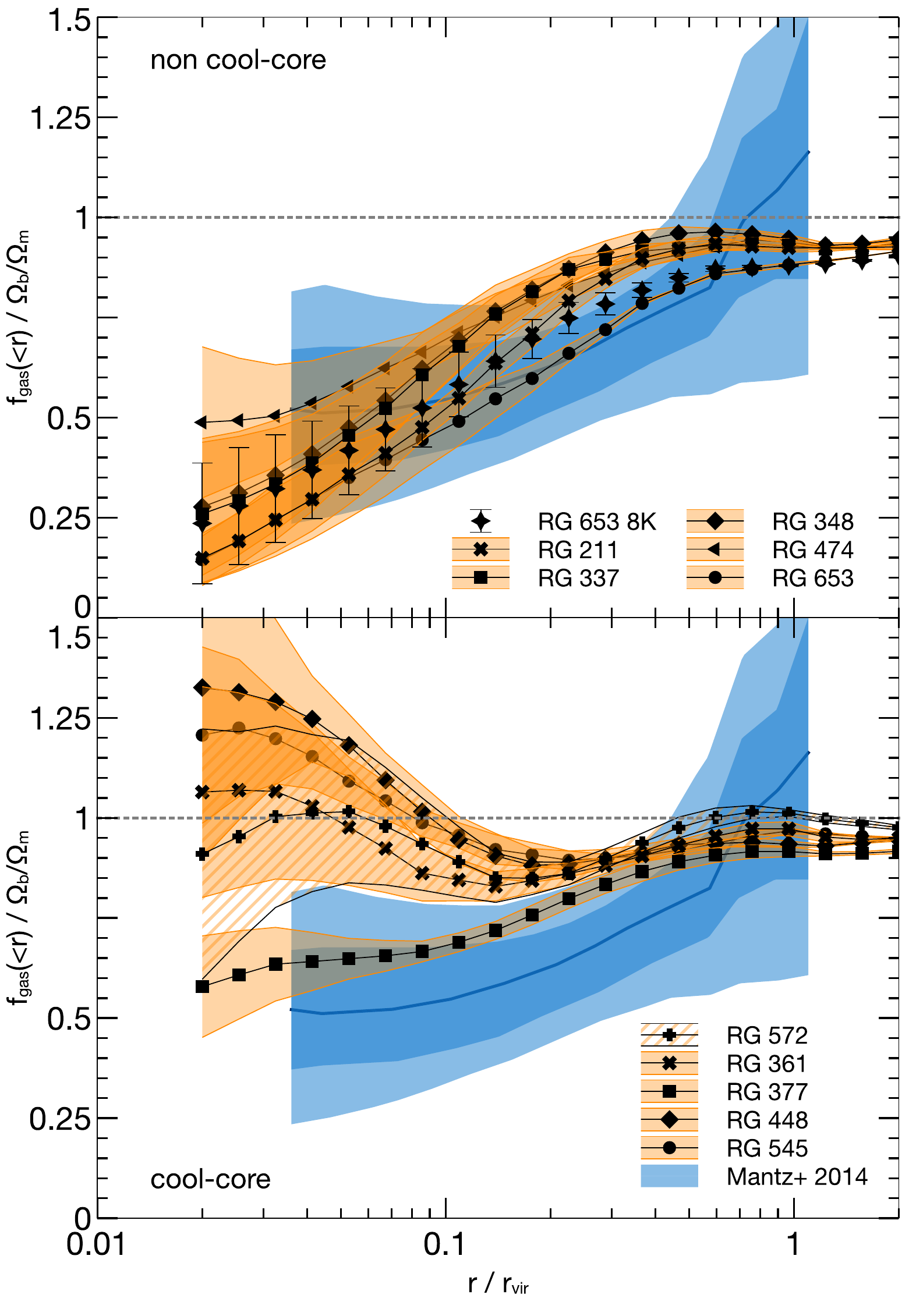}
\end{center}
\caption{\label{fig:gas_depletion}Comparison of cumulative gas fraction profiles for the same stacked clusters as in Figure~\ref{fig:accept_profiles}, i.e. for the mass range $4<M_{500c}/10^{14}M_\odot < 6$ with the observational results from \citet{Mantz:2014}. We split again into cool core (bottom) and non-cool-core clusters (top), and show stacked results for each simulated cluster individually. The shaded ribbons indicate the $1\sigma$ in the stack of each cluster. In the bottom panel, we have indicated the fossil cluster RG~572 with a hatched ribbon.}
\end{figure}

A crucial observable reflecting the degree to which collisional matter follows the total mass distribution of galaxy clusters is given by the gas depletion profiles. Especially in the cluster cores, cooling and AGN feedback will affect both the gas and the dark matter distribution. In particular, if the central black hole accretes gas quasi-periodically, AGN feedback can transform the cuspy dark matter profiles into cored profiles \citep[][consistent with what we find in the RHAPSODY-G simulations]{Martizzi:2013}. In Figure~\ref{fig:gas_depletion}, we show the ratio of enclosed gas mass to total mass in units of the universal baryon fraction. We split our sample of clusters into non-cool-core (top) and cool-core (bottom) systems. We compare our results to the observational relations of \cite{Mantz:2014}, as indicated by the blue ribbons. 

The NCC systems follow the observational relation reasonably well. The fossil cluster and the CC systems however show a very large central gas fraction, in many cases even above the universal gas fraction \cite[similar to the results of][]{Burns:2008}. This result is clearly in tension with the results of \cite{Mantz:2014}, and reflects the high electron densities and very low central entropies we have seen for the CC systems above. The conclusion must be that in the case of the CC systems, the AGN feedback is not able to stabilise the core at levels consistent with X-ray observations. The exciting possibility is that other forms of non-thermal feedback (or processes) must be plausibly involved in order to bring these results in line with observations. We note that our results for NCC systems are consistent with published results from SPH simulations \citep[e.g.][]{Battaglia:2013,Sembolini:2013,Planelles:2013} at scales of $~0.1\,R_{\rm vir}$, but are somewhat higher at larger radii.

Less visible but as significant is the discrepancy of the gas fraction at $\sim {\rm R}_{\rm vir}/2$, where the profiles are slightly but systematically high with very little scatter. As we see below in Section~\ref{fig:fgas_AGN}, the thermal AGN feedback does not reach these large radii in our simulations, and it is not possible to tune the energy injection parameter $\Delta T$ to deplete the cool cores efficiently.

Despite these discrepancies, the use of the gas fraction at $R_{2500}$ as a robust cosmological observable \citep{Allen:2008,Mantz:2014} is strongly supported by our simulations since $R_{2500}$ appears to be outside the reach of the AGN and shows a virtually unbiased thermal mass (c.f. Section~\ref{sec:hydrostaticmass}).

\begin{figure}
\begin{center}
\includegraphics[width=0.8\columnwidth]{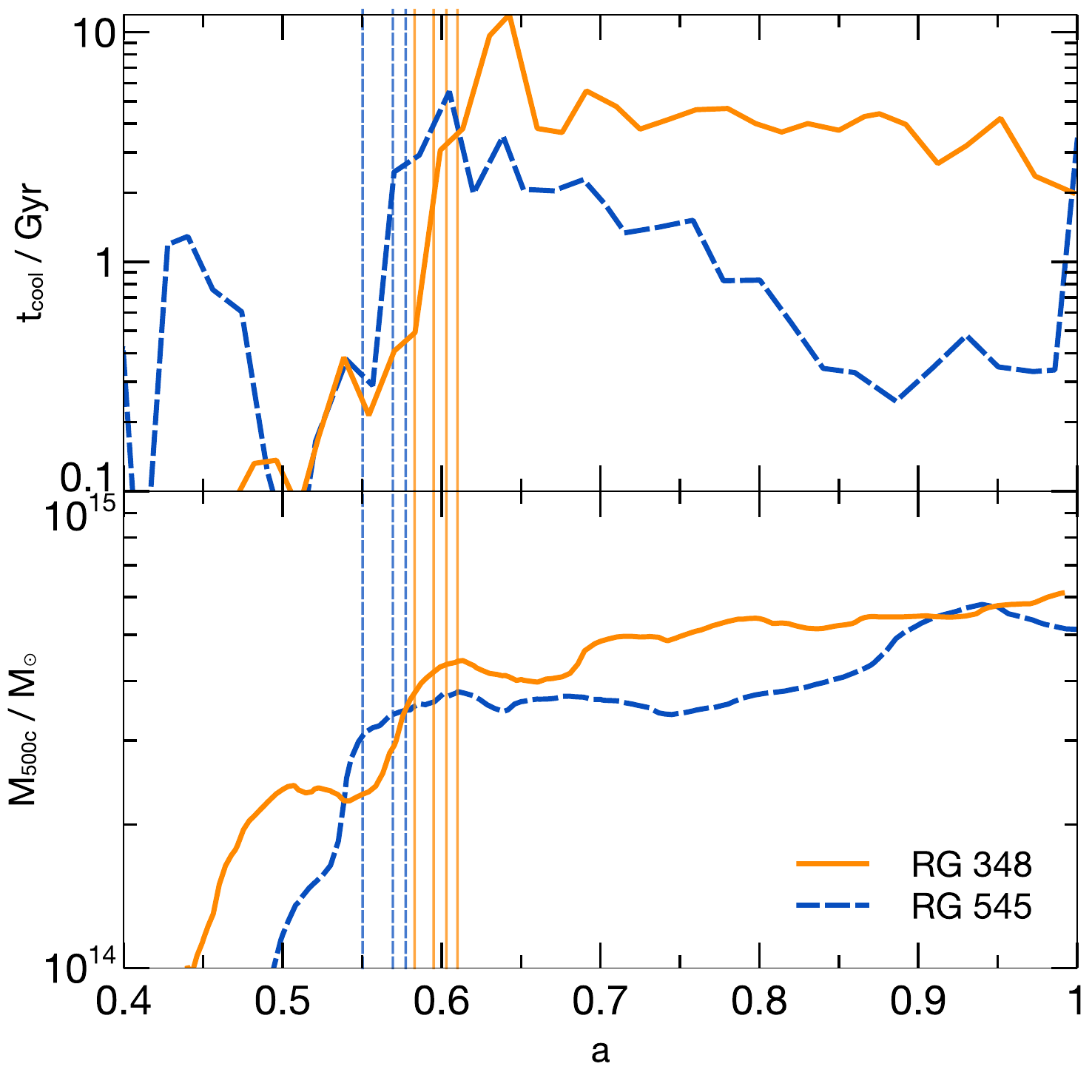}
\end{center}
\caption{\label{fig:assembly348}Assembly history of RG~348 and RG~545.  The first is a non-cool-core cluster at $z=0$, the latter is a cool core cluster. The top panel shows the cooling time evolution in the core for the two systems, the bottom panel the mass accretion history. Both undergo a major merger between $a\sim0.5$ and $a\sim0.6$. While the core of RG~348 is substantially heated during the merger, the effect on RG~545 is not as dramatic. The vertical lines indicate the times for which we show images illustrating the evolution of RG~545 and RG~348 in Figure~\ref{fig:merger545} and \ref{fig:merger348}, respectively.}
\end{figure}

\begin{figure}
\begin{center}
\includegraphics[width=0.95\columnwidth]{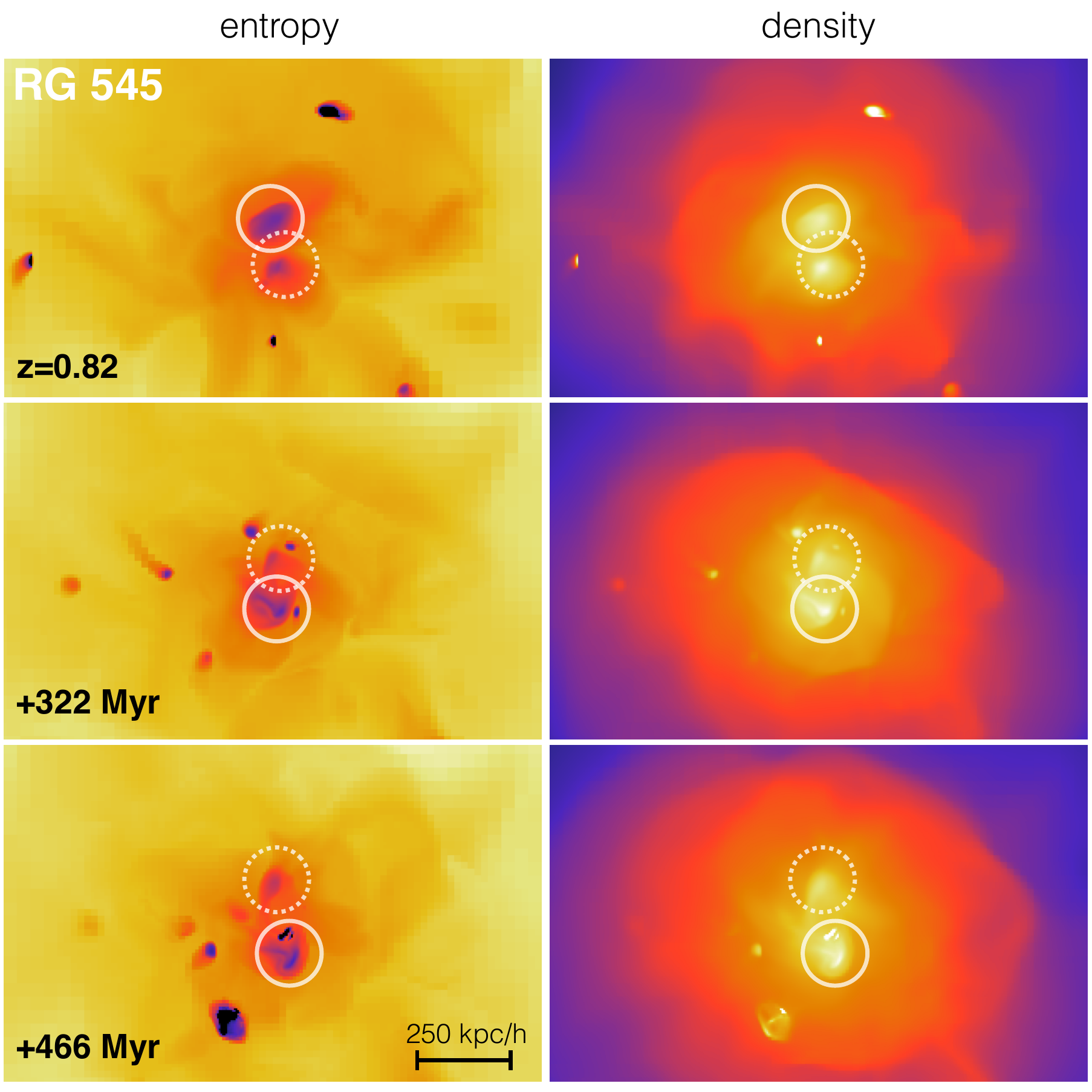}
\end{center}
\caption{\label{fig:merger545}Cool core surviving after a major merger (RG~545): We show the entropy (left column) and density (right column). The first row is at $z=0.82$, the next rows are 322 and 466~Myr later. The top left panel clearly shows the two cool cores of the systems before the merger, but due to enough angular momentum, the cores do not collide (unlike in the case of RG~348, see Figure~\ref{fig:merger348}), and the cool core survives. Solid and dashed circles indicate the cores of the main cluster and the merging cluster, respectively.}
\end{figure}

\begin{figure*}
\begin{center}
\includegraphics[width=2\columnwidth]{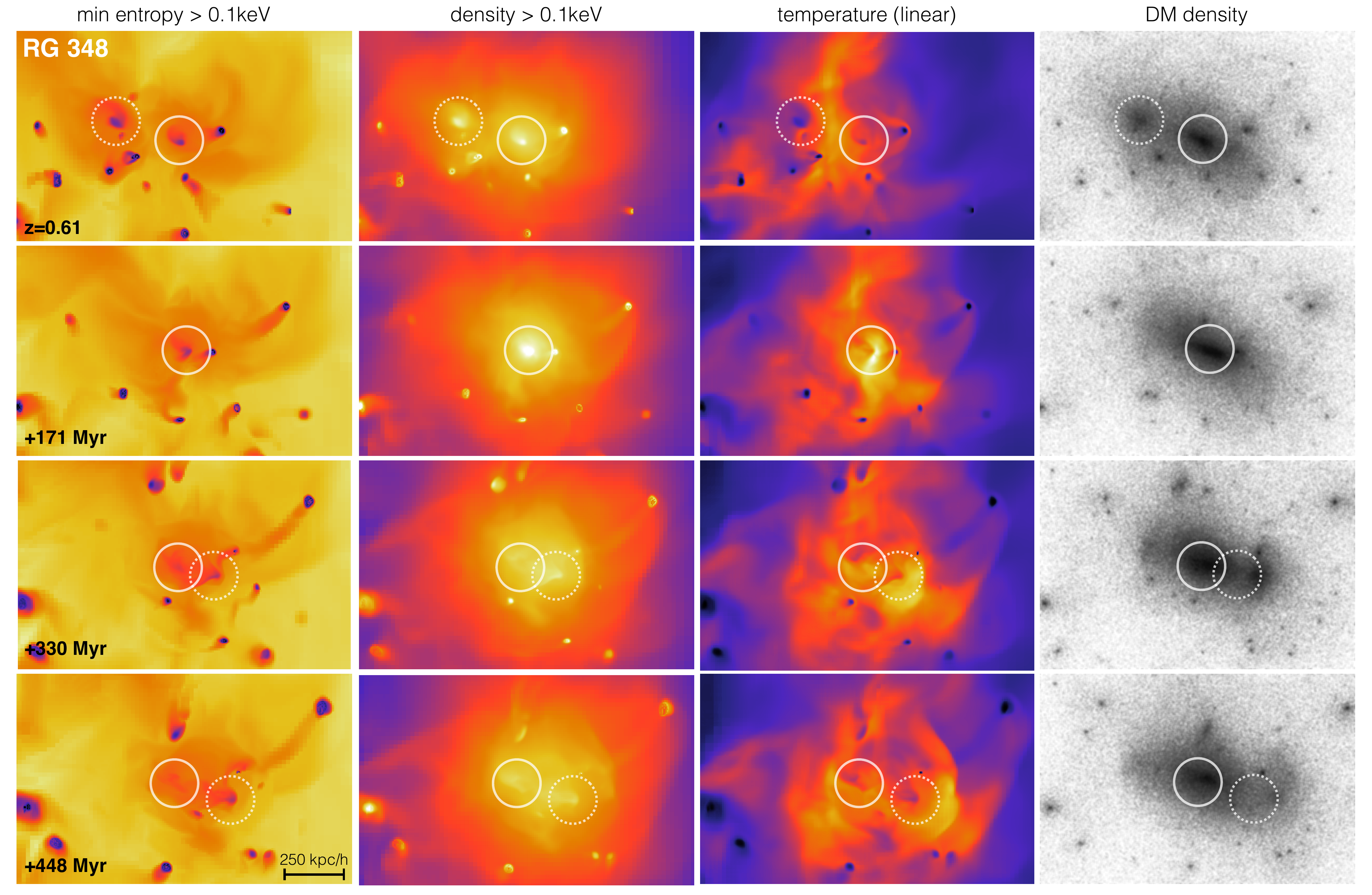}
\end{center}
\caption{\label{fig:merger348}Cool core destroyed in a major merger (RG~348). We show the minimum entropy along the line-of-sight (left column) and the density (middle left) of gas with a temperature of at least 0.1~keV, as well as the density weighted temperature in linear scale (middle right) and the dark matter density (right column). The first row is at $z=0.61$, the next rows are 171, 330, and 448~Myr later. The top left panel clearly shows the two cool cores of the systems before the merger.  During the merger, the cores undergo a strong shock (clearly visible in 2nd temperature panel from top). After the collision, the cool cores are dissipated and undergo mixing with higher entropy gas. Solid and dashed circles indicate the cores of the main cluster and the merging cluster, respectively. The many low entropy blobs are satellite galaxies that do not carry much mass with them. The depth of the images is comoving $1\,h^{-1}{\rm Mpc}$ centred on the more massive cluster.}
\end{figure*}


\section{The origin of the cool-core / non-cool-core dichotomy}
\label{sec:origin_NCC}
We next inspect the origin of the CC/NCC dichotomy in our simulations. We find that at early times, $z\sim0.6$ and $z\sim0.8$, respectively, both RG~348 and RG~545 are cool-core clusters and have a very similar assembly history, which we show in Figure~\ref{fig:assembly348}. Both undergo a major merger with a similar mass ratio at similar times. However, the cooling time of the core (top panel of Figure~\ref{fig:assembly348}) rises dramatically more in the case of RG~348 to a value of about 5~Gyr after an initial higher spike. Cluster RG~545 only experiences an increase in cooling time to $\lesssim2.5\,{\rm Gyr}$.  We define the core here as the gas enclosed in an overdensity of $8\times10^4$ and evaluate the cooling time at that radius (we find that the free-fall time reduces only slightly over the same time scale so that roughly $t_{\rm cool}/t_{\rm ff}\propto t_{\rm cool}$). After the merger, both clusters have a more quiescent merger history and both cores start to cool again. The cooling of RG~545 is much more rapid, plausibly since it is quickly experiencing runaway cooling, while RG~348 cools at a slower rate --- the time to $z=0$ is still 7~Gyr, shorter than its cooling time --- most likely reflecting AGN heating that is able to keep the core hot.

After inspecting the time evolution of the systems, the difference in how the mergers proceed is obvious. In the case of RG~545, see Figure~\ref{fig:merger545}, the cores do not collide since the merger has large angular momentum, so that the cool core is perturbed but not destroyed and is just sloshing in the centre of the halo. Quite different is the merger that RG~348 experiences: here the two cool cores of the progenitor clusters collide and dissipate within $\sim100\,{\rm Myr}$. We show their time evolution over $\sim450\,{\rm Myr}$ during the merger in Figure~\ref{fig:merger348}: the left column shows the minimum entropy along the comoving $1\,h^{-1}{\rm Mpc}$ image depth of X-ray emitting gas with a temperature above 0.1~keV. We also show the density of gas above 0.1~keV, as well as the temperature (in linear scale) and the dark matter density. For RG~545, we only show entropy and density since the merger is less spectacular. From the sequence of RG~348, one sees that the cores pass right through each other (in the moment shown in the second row); after the collision, the cool core of the main cluster halo is destroyed, while the one of the less massive progenitor survives weakened. In the last time shown and subsequently, the surviving core mixes quickly with the hot cluster gas due to ram pressure and Kelvin-Helmholtz instabilities. We also note that, as expected, we detect a weak displacement between the BCG as well as the dark matter halo centre and the gas centre after the collision, consistent also with the observational findings of e.g. \cite{Semler:2012}.  

\begin{figure*}
\begin{center}
\includegraphics[width=2.1\columnwidth]{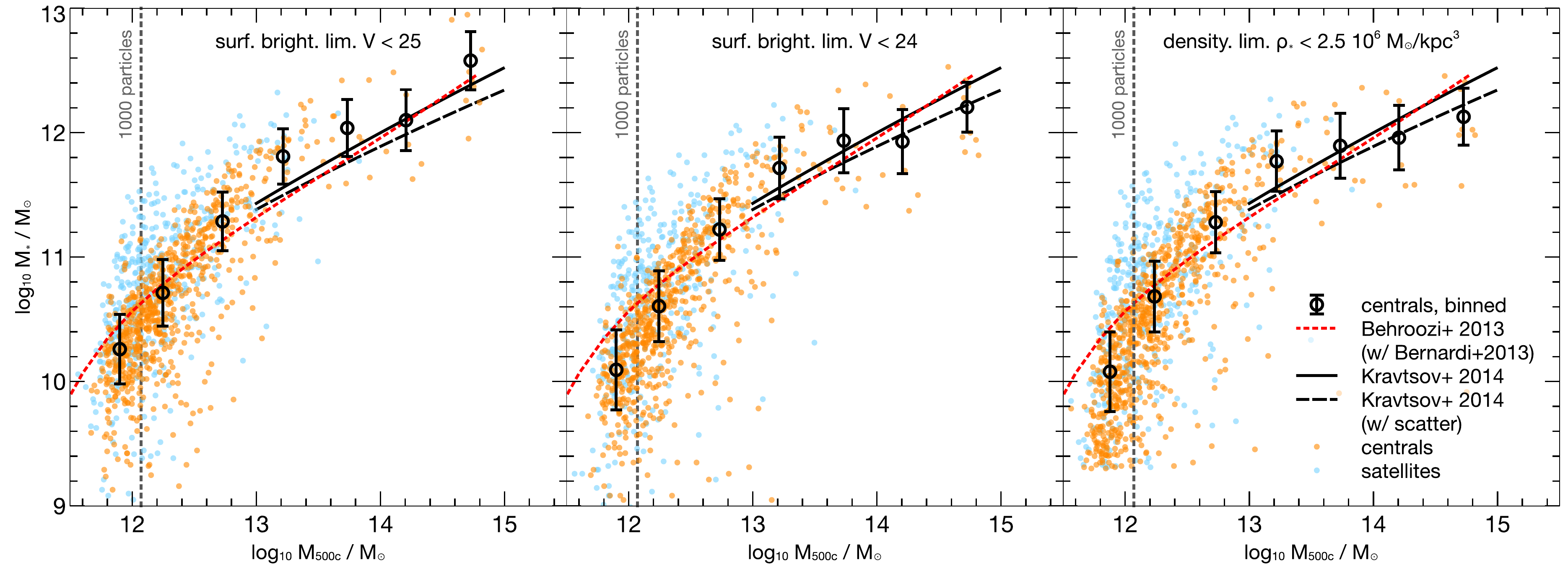}
\end{center}
\caption{\label{fig:mhalo_mstar}
Comparison between the $M_\ast$-$M_{500c}$ relation for galaxies from all 10 {\sc Rhapsody}-G haloes at $z=0$ and abundance matching constraints. The orange dots represent central galaxies with the mean stellar mass in bins of halo mass shown by bold black circles with errorbars indicating the 1$\sigma$ scatter, while the blue dots represent the satellite population, respectively. The solid and dashed black lines show the abundance matching results from \citet{Kravtsov2014} using the stellar mass functions from \citet{Bernardi2013}, while the gray shaded area shows the abundance matching results from \citet{Behroozi2013} (remapped from $M_{200c}$ to $M_{500c}$ using the mean relation from the simulation itself). Simulated satellites appear at higher stellar masses as a result of a reduction of their halo mass due to tidal stripping. The drop of stellar masses below the Behroozi relation for masses $\lesssim10^{12}\,{\rm M}_\odot$ is a resolution effect (c.f. Fig.~\ref{fig:mhalo_mstar_8k}). }
\end{figure*}

The finding of a connection between CC/NCC and mergers in early simulations put forward by \cite{Burns:2008} and \cite{Planelles:2009} thus should be complemented by this new finding.  Not only slowly accreting systems can have cool cores; the amount of angular momentum in major mergers (i.e. whether the collision is head-on in such a way that the cores do pass through each other) appears to be of high importance as well. This is also consistent with the findings from idealised non-cosmological cluster merger simulations by \cite{Poole:2008}. {\em Major mergers are thus a necessary but not sufficient condition for the destruction of cool cores}. The survival of RG~545 as a CC system clearly shows that the cores are not easily disrupted, in some cases, even by a major merger. For all clusters in our sample it is true that, very similar to the two examples in Figure~\ref{fig:assembly348}, the cool core clusters never had their cooling time increased beyond at most 20 times the core free fall time. This is roughly consistent with \cite{Meece:2015} who find that a ratio of at most 10 is needed in order to undergo a significant thermal instability. Since, absent other heat sources, every cluster will slowly cool again after a major merger event (and basically every cluster will experience one or more major mergers throughout its history), the sharp distinction between the CC and NCC systems in terms of their profiles then arises simply due to how much a merger perturbs the core --- determined by the merger's angular momentum --- and whether or not rapid cooling can set in subsequently. In the case of RG~348, the cool core has been dissipated in no more than a few 100~Myr with a subsequent cooling time that is significantly higher. {\em The disparity in these two time-scales makes this a quick transition, allowing the bimodality apparent in the cluster population to be sustained.} 



\section{Properties of the galaxies in and around the simulated clusters}
\label{sec:stellar_properties}
We next discuss the properties of the galaxy population in and around our simulated clusters. Since the high-resolution region is an $8\,h^{-1}{\rm Mpc}$ sphere around each cluster at $z=0$, we have a sizeable sample of lower mass haloes ranging from lower mass clusters to group to galaxy scales in the outskirts of our massive clusters. This facilitates a statistical comparison with the more field-dominated observational samples. Nevertheless all galaxies in our simulations do live in a high-density environment, which should be kept in mind when comparing these results to observations.


\subsection{Comparison with abundance matching results}
\label{sec:abundance_match}
Apart from the hot intracluster plasma, galaxy clusters harbour a large population of satellite galaxies. A general short-coming of galaxy cluster simulations is that in these simulations it is computationally not affordable to resolve the length scales relevant to important aspects of galaxy formation (on say 100~pc and smaller). The hope is that at least the most massive galaxies are resolved well enough. A bona-fide test of the realism of our simulations is thus to investigate whether the central cluster galaxies indeed have stellar masses and other properties that are consistent with observations. Remarkably, despite the fairly poor resolution at galaxy mass halo scales, we find that the star formation efficiency and supernova feedback as dictated by the subgrid models is in fact able to reproduce accurate stellar masses across all halo masses that we resolve with $\gtrsim1000$ particles in our simulations. In \cite{Wu:2015}, based on the same simulations discussed here, we find stellar mass fractions of $\sim 10\%$ of the baryon fraction, consistent with observational constraints.

In Figure~\ref{fig:mhalo_mstar}, we show the comparison of the stellar masses of our galaxies (satellite and central) as a function of their halo mass with results obtained using the abundance matching technique \citep[c.f. e.g.][]{Conroy2006,Moster:2010,Behroozi2013}. The very massive systems that we consider here occupy however the very tail of the stellar mass functions, and in fact, the older abundance matching relations \citep[e.g.][]{Moster:2010,Behroozi2013} do not capture very well the scaling at the highest masses due to the way stellar mass was counted in the stellar mass functions used. More accurate results have been obtained by \cite{Kravtsov2014} taking explicit care of the intracluster light component; these are fully consistent with the the relations of \cite{Behroozi2013} when updated with the stellar mass functions of \cite{Bernardi2013}. We show a comparison of these relations with the stellar masses from our simulations measured using a range of techniques. In particular, we have produced mock V-band images and count the stellar mass inside of isophotes where the surface brightness is above $25$ and $24\,{\rm mag}\,{\rm arcsec^2}$, as well as using a simple spherical density threshold of $2.5\times10^6\,{\rm M}_\odot/{\rm kpc}^3$ (panels from left to right in Figure~\ref{fig:mhalo_mstar}).

\begin{figure}
\begin{center}
\includegraphics[width=0.9\columnwidth]{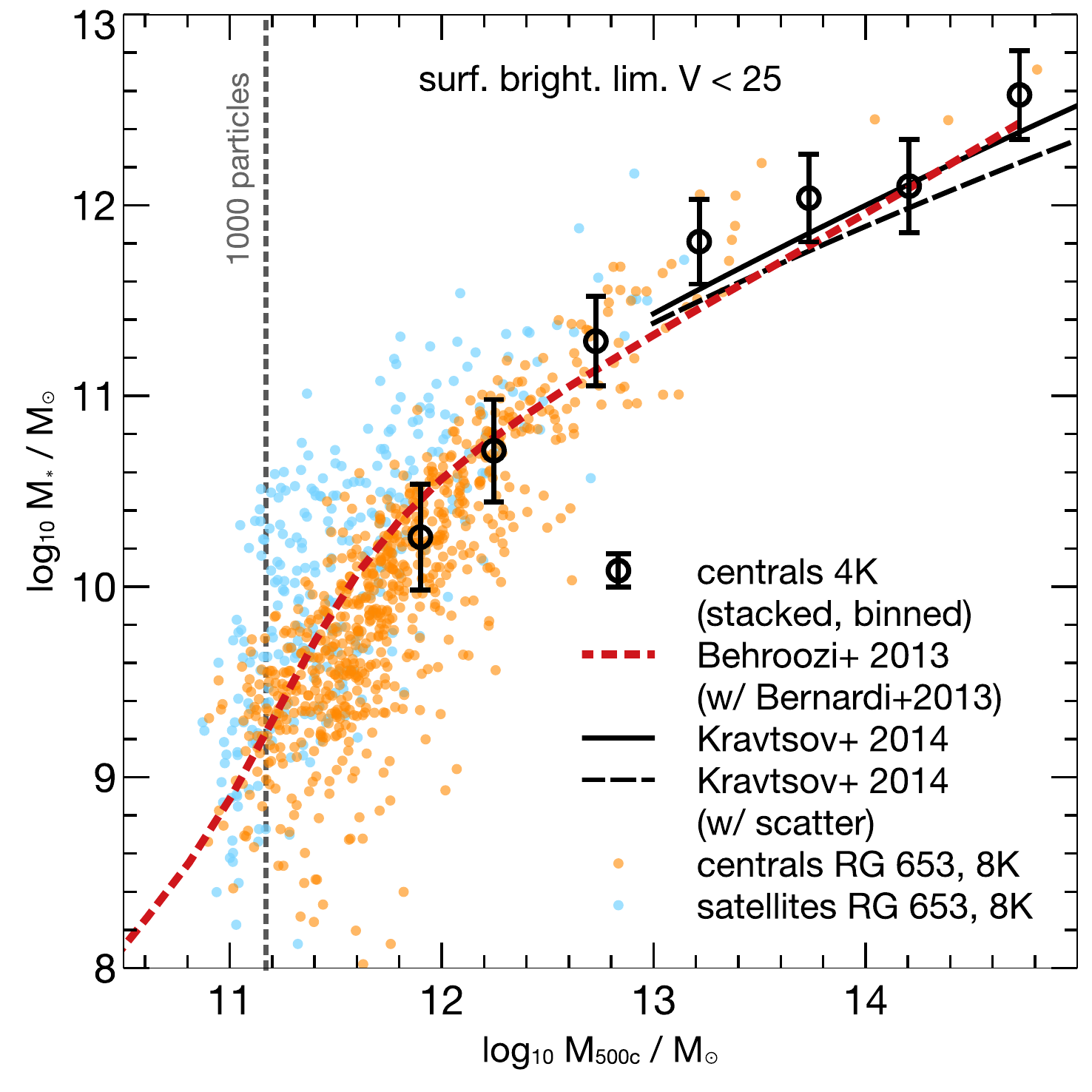}
\end{center}
\caption{\label{fig:mhalo_mstar_8k}As Figure~\ref{fig:mhalo_mstar}, but a comparison between the $M_\ast$-$M_{500c}$ relation for galaxies from a higher-resolution simulation of cluster RG~653 (simulated at 8 times better mass and twice better spatial resolution than the full sample) at $z=0$ and abundance matching constraints. }
\end{figure}

We see that, while we largely overproduce stellar mass compared the original \cite{Behroozi2013} result, our simulations are consistent with more recent estimates for massive systems.  The stellar masses in massive haloes above $M_{\rm 500c}\sim 2\times10^{12}\,{\rm M_\odot}$ are still somewhat high, but not dramatically so, and details depend on the exact definition of how stellar mass of central group and cluster galaxies should be measured. Notably, the stellar masses at galactic halo scales seem to undershoot the observational relations. However, this is a resolution effect. In Figure~\ref{fig:mhalo_mstar_8k}, we show the respective plot for the galaxies in and around cluster RG~653 simulated at twice better spatial and eight times better mass resolution. With higher resolution, the full range of halo masses from $10^{11}-10^{15}\,{\rm M}_\odot$ is in good agreement, with the slight overproduction of stars in the most massive haloes remaining.

\subsection{Star formation rates}

\begin{figure}
\begin{center}
\includegraphics[width=0.9\columnwidth]{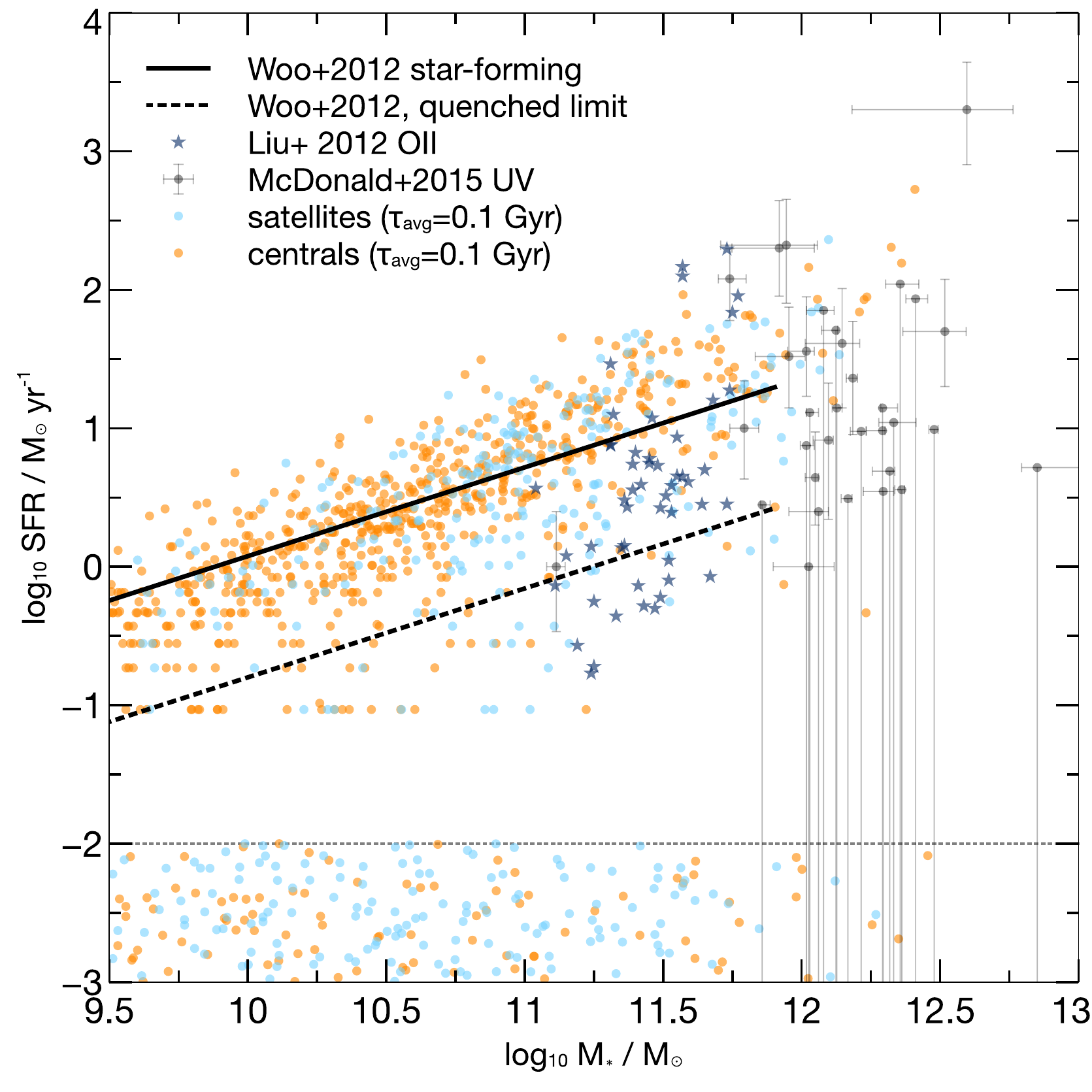}
\end{center}
\caption{\label{fig:mstar_sfr}Star formation rate as a function of stellar mass for central (orange) and satellite (blue) galaxies from all 10 {\sc Rhapsody}-G haloes at $z=0$. Galaxies with no detectable star formation in the previous 500 Myr have been assigned random values between $10^{-3}$ and $10^{-2}\,{\rm M}_\odot/{\rm yr}$. For comparison, the main sequence of star forming galaxies in SDSS from \citet{Woo2012} (solid black line) and the division between star-forming and quenched galaxies (black dotted) along with the OII-derived star formation rates for the BCG samples of \citet{Liu2012} and the UV estimates of \citet{McDonald:2015} are shown (note that for the latter, data points with only long lower error bars are upper limits).  }
\end{figure}

We next want to investigate the reason for the somewhat high stellar masses in the most massive haloes in our simulation. To this end, we show each of our simulated galaxies (for all clusters) in the stellar-mass vs. star formation rate plane. We estimate our star formation rates as the mass of star particles in each simulated galaxy that have an age of less than $100\,{\rm Myr}$ divided by the $100\,{\rm Myr}$. Galaxies which do not have any star particle fulfilling this condition get assigned a random star formation rate (SFR) between $10^{-2}$ and $10^{-3}\,{\rm M}_\odot\,{\rm yr}^{-1}$. We perform this step in order to allow a visual estimate of the fraction of quenched galaxies in Figure~\ref{fig:mstar_sfr}.

From Figure~\ref{fig:mstar_sfr}, we see that our galaxies with significant star formation follow a narrow star forming sequence that is consistent in normalisation and slope with the SDSS star forming sequence as measured by \cite{Woo2012} (solid black line). In addition, we show the star formation rates determined by \cite{Liu2012} for a sample of central cluster galaxies from SDSS, and those of \cite{McDonald:2015} for BCGs of SPT clusters. When comparing with the limit below which \cite{Woo2012} would classify galaxies as ``quenched'' (black dashed line), we see that many more massive galaxies in this simulation are not quenched --- the SDSS quenched fraction exceeds 50 per cent above $M_\ast\sim10^{10.5}$ in \cite{Woo2012} (private communication from J. Woo). While the SDSS results are averaged over all field galaxies, it may be that some biases exist in these most dense environments around the most massive clusters. However, it seems more plausible that these simulations either miss an additional physical mechanism or lack the resolution to reproduce a realistic quenched fraction at the intermediate masses. While the highest star formation rates we observe are roughly consistent with \cite{Liu2012} and perfectly consistent with \cite{McDonald:2015} (note that the data points with one-sided lower error bar are upper limits only), it is clear that quenching of intermediate (and maybe even high) mass galaxies is inefficient in our simulations. This is consistent with the somewhat high stellar masses compared to the abundance matching results. Finding the mechanism(s) necessary to suppress star formation in massive galaxies is among the most pressing tasks of galaxy formation theory today and it is not surprising that our simulations are not performing significantly better than better resolved simulations dedicated to study galaxy formation.

\section{Implications for cluster cosmology: Evolution along scaling relations}
\label{sec:scalingrelations}

The use of cluster counts as a cosmological probe exploits population scaling relations that link total system mass to observable properties of the galaxies or hot plasma.  In this section, we compare the evolution of the simulated clusters along a range of scaling relations with observational results from X-ray and CMB data and some other published results from simulations. We focus on the Sunyaev-Zeldovich effect as well as on X-ray luminosity and temperature. In the absence of a large statistical sample of clusters, tracking a few clusters over time allows us to relate the role of evolutionary processes to their impact on observables. 

Since systems in the simulated sample have roughly the same mass at $z=0$, their evolutionary tracks can be used to assess the contribution of their assembly history to scatter in the mass-proxies. In order to study the evolution of the scatter or additional biases relevant to the full population, however, larger samples would be required.   While we compare our scalings to observations, we caution that this type of exercise is non-ideal in that: i) we employ true, three-dimensional spherical masses in simulations while observational estimates may be biased with respect to these values;  ii) similarly, the intrinsic properties of the simulations are measured directly in the simulations rather than derived from models applied to mock observations, and; iii) the statistics of the simulated sample are not necessarily representative of a broader mass-complete sample.   The purpose of this exercise is to identify areas of agreement and discrepancy, which in turn should provide insights into future adjustments to the astrophysical feedback model.  


\subsection{Sunyaev-Zeldovich masses}
\label{sec:szmasses}
\begin{figure}
\begin{center}
\includegraphics[width=0.9\columnwidth]{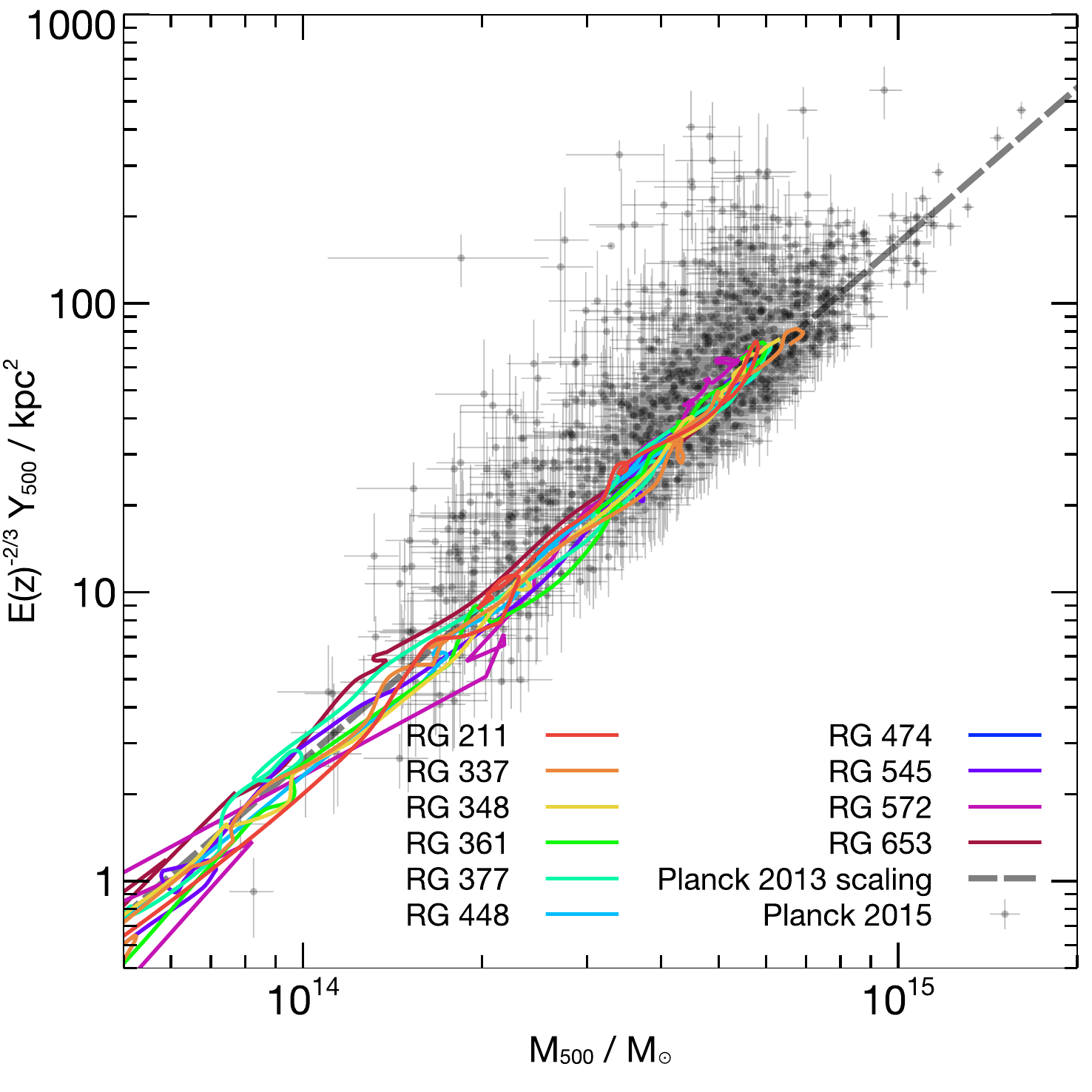}
\end{center}
\caption{\label{fig:MSZ}Comparison of the {\em Planck} 2015 SZ cluster sample with the evolutionary tracks of the RG clusters along the M-Y relation. The {\em Planck} data is taken from the 2015 union catalog with rescaled $Y_{500}=Y_{5R500}/1.814$  according to \citet{Melin:2011}.}
\end{figure}

We first compare the  Compton-$y$ parameters integrated over our simulated clusters, i.e. calculated as
\begin{equation}
Y_{500} = \int_{|\mathbf{x}|<R_{500}} \sigma_T \frac{k T}{m_e c^2} n_e\,{\rm d}^3 x,
\end{equation}
with SZ masses and Y5R500 measurements from the {\em Planck} HFI union catalog 2015 (R2.08), see \cite{PlanckXXXII:2015}. We have converted the {\em Planck} Y5R500 values to $Y_{500}$ using the ratio of 1.814 between the measurement at 5 and at 1$R_{500}$ quoted in \cite{Melin:2011} and assumed the {\em Planck} cosmological parameters when converting Y5R500 to units of ${\rm kpc}^2$. We show the results of this comparison in Figure~\ref{fig:MSZ}, along with the unbiased (i.e. setting the hydrostatic mass bias to $b=0$) {\sc Planck2013} mean baseline relation 
\begin{equation}
E^{-2/3}(z) \left[\frac{Y_{500}}{{\rm kpc}^2}\right] = 10^{1.81\pm0.02} \left[ \frac{M_{500}}{6\times 10^{14} {\rm M}_\odot} \right]^{1.79\pm0.08}
\end{equation}
as quoted in \cite{Planck2013_XX}. 

\subsection{X-ray observables}
\label{sec:xrayobservables}
\begin{figure}
\begin{center}
\includegraphics[width=0.9\columnwidth]{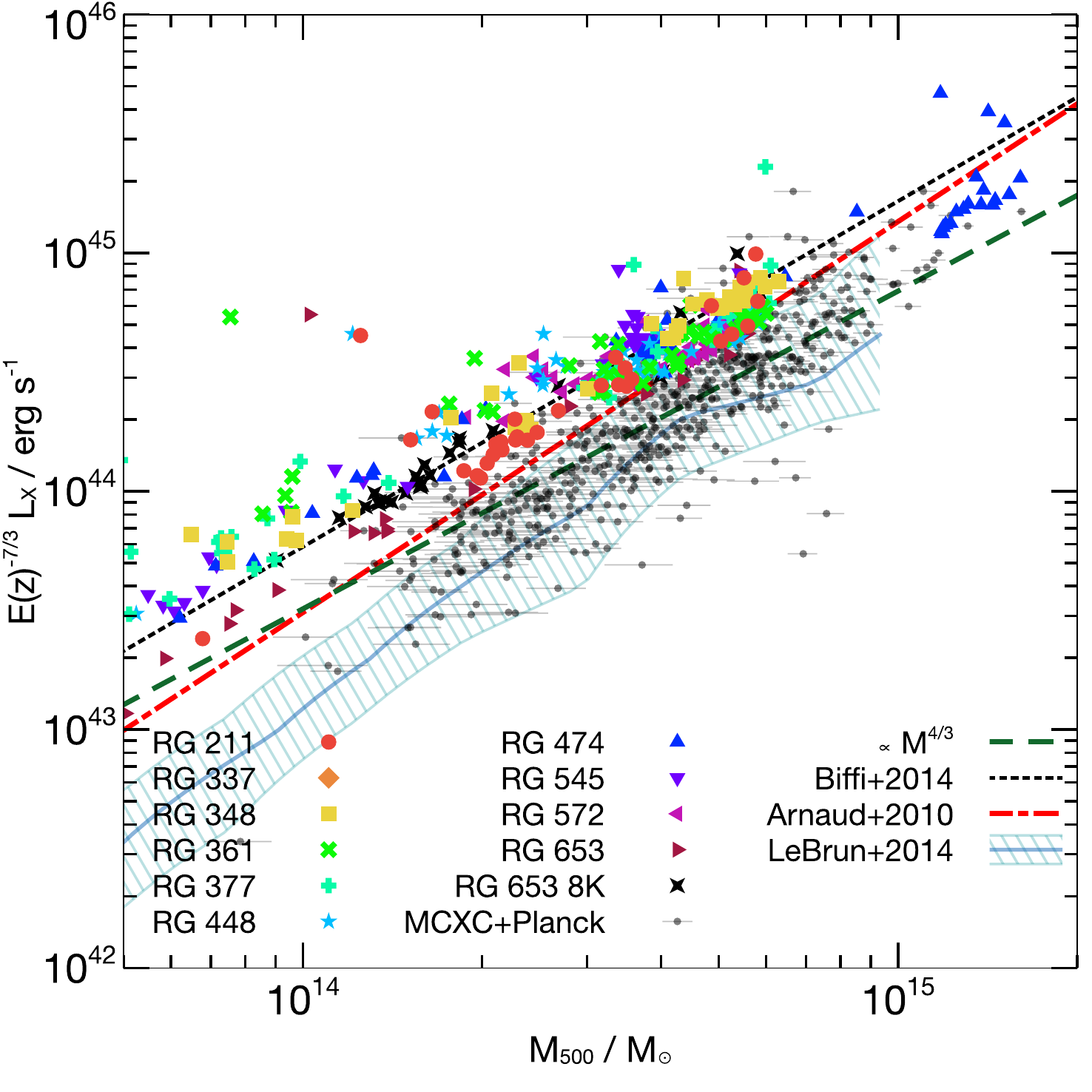}
\end{center}
\caption{\label{fig:MCXC}Comparison of the $M_{500}-L_X$ scaling relation with observations and other simulations. The RG clusters are again shown as evolutionary tracks, the luminosity is the integrated 0.1-2~keV luminosity estimated from APEC spectra. For the clusters from the MCXC sample we have used the 0.1-2~keV luminosities from MCXC {\em but the respective Planck SZ masses} here. The dashed-dotted line represents the fit from \citet{Arnaud:2010}, which is almost identical to the scaling relation used in MCXC, and the dashed line is the best fit self-similar solution (i.e. the $\propto M_{500}^{4/3}$ fit to the MCXC/Planck joint data). The dotted line shows the best fit from \citet{Biffi:2014}, who however use the wider Chandra band, while the blue hatched region is from \citet{LeBrun:2014}, and uses also the 0.1-2 keV band.}
\end{figure}

\begin{figure*}
\begin{center}
\includegraphics[width=2.1\columnwidth]{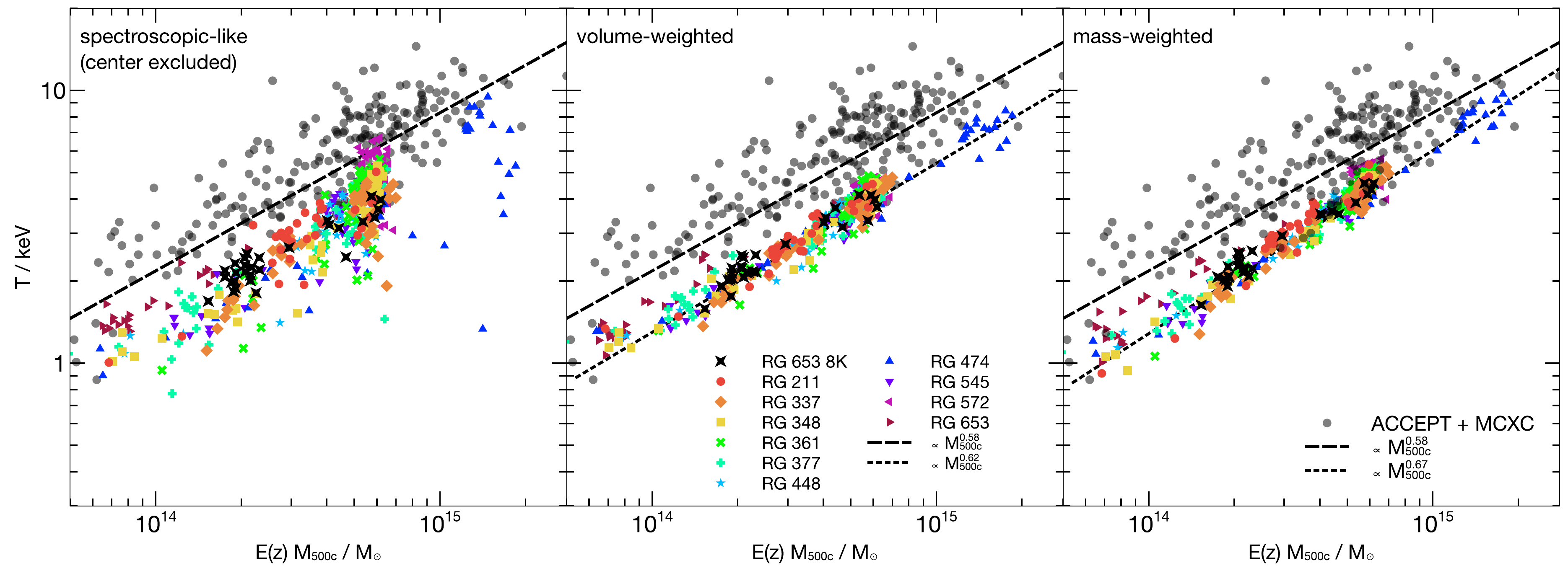}
\end{center}
\caption{\label{fig:MTX}Comparison of the ACCEPT/MCXC clusters $M500-T_X$ relation with the evolutionary tracks of the RG clusters. From left to right, we show comparisons with different estimates of the gas temperature in the simulations:  ({\em Left}) the spectroscopic-like temperature, with core excluded, ({\em Center}) volume-weighted temperature and ({\em Right}) the mass weighted temperature. The gray dots are a combination of X-ray masses from the MCXC catalog with temperature estimates taken from the ACCEPT catalog, and the black long-dashed line is the best fit power law to this data, while the short-dashed line indicates a power-law fit to RG~474, which covers the largest range in mass evolution.}
\end{figure*}

Our simulations show good agreement with the {\sc Planck2013} baseline for all systems over the entire range from $\sim5\times10^{13}$ to $10^{15}\,{\rm M}_\odot$. Despite the large variance between the clusters' assembly histories, we find very little scatter, $\sim0.2$~dex around the {\sc Planck2013} mean relation \citep[consistent with other simulations, e.g.][]{Sembolini:2013}, with only the fossil cluster RG~572 being somewhat of a significant outlier (though still insignificant given the scatter in the {\em Planck} $M-Y$ data). In fact, it is remarkable that RG~572 does {\em not} stand out. In particular, the scatter is substantially smaller than the scatter around the $M-Y$ scaling relation in the {\sc Planck2015} SZ catalog.  The large observed scatter is mostly due to complications from projection and low angular resolution, which requires joint fitting of $Y_{5R500}$ and cluster size $\theta_{500}$.  The large and markedly asymmetric scatter in the Planck $M-Y$ data is an artefact of our simplistic rescaling from $Y_{5R500}$ to $Y_{500}$ with a constant factor and appears in rather stark contrast to the simulation results.  Investigating the simulation prediction for $Y_{5R500}$ would require a matched filter approach to reduce interloper effects at larger radii, as employed by the {\it Planck} team, and possibly full lightcone projections \citep[e.g.,][]{White2002}, which are beyond the scope of this work.    


In Figure~\ref{fig:MCXC}, we show a comparison of the X-ray luminosity of the RG clusters during their evolution as a function of their mass with various scaling relations from the literature as well a direct comparison with clusters from the MCXC sample \citep{Piffaretti:2011} using the Planck SZ masses and the MCXC luminosity (since otherwise one simply reproduces the scaling relation used to estimate the masses, which is very close to \citealt{Arnaud:2010}). We measure the X-ray luminosity by computing the emissivity per cell using the APEC tabulated emission model and exclude the central core region inside $0.15\,R_{500}$, i.e.
\begin{equation}
L_{\rm X} = \int_{0.15<|\mathbf{x}|/R_{500}<1} k T \epsilon(T,Z)\,n_e\,n_{\rm H}\,{\rm d}^3 x,
\end{equation}
where $n_e(\mathbf{x})$ and $n_{\rm H}(\mathbf{x})$ are the electron and hydrogen number density, $T(\mathbf{x})$ is the electron temperature, and $\epsilon(T,Z)$ is the (temperature- and metallicity-dependent) emissivity taken from the APEC tables and integrated over the same energy range $0.1-2\,{\rm keV}$ as the MCXC clusters. The exclusion of the core is necessary as it introduces a strong variability of the X-ray luminosity due to AGN events where the central density fluctuates and unrealistically high luminosities can be obtained. The samples that we compare against however do include the core, so that the comparisons have to be taken with a grain of salt. For comparison, we also show the results of \cite{LeBrun:2014}, who use the same AGN model in SPH simulations and whose X-ray luminosity appears to be somewhat low compared to the MCXC sample. For completeness, we also show the best-fit relation from \cite{Arnaud:2010}, which is basically identical to the scaling relation used for the $M_{500}$ measurements in MCXC, as well as the best fit from \cite{Biffi:2014}. The latter authors use the Chandra bolometric band ($0.3-12\,{\rm keV}$), which explains the higher X-ray luminosity compared to the MCXC data. In comparison, our clusters evolve along a scaling relation that is somewhat shallower than both \cite{Arnaud:2010} and \cite{LeBrun:2014},  consistent with \cite{Biffi:2014} (who however used the larger spectral window) but slightly steeper than a simple self-similar scaling $\propto M^{4/3}$ (shown as a green dashed line in the figure and whose normalisation has been obtained by fitting to the MCXC data). In fact, the best-fit slope in our case is closer to $1.2$, but we do not want to make a more quantitative statement at this stage since the comparison of evolving small (autocorrelated) samples should be taken with some caution for making rigorous predictions. It suffices to observe here that the simulation X-ray luminosities are consistently higher (by about a factor of two around $10^{14}M_\odot$ and at most $\sim 20$ per cent at $10^{15}M_\odot$) than the MCXC luminosities, with a scaling relation that is slightly steeper than the self-similar expectation of $4/3$. There is a weak indication that the normalisation of the evolutionary track of clusters is persistent, i.e. RG~211 tends to be lower than RG~361, which is lower than RG~348 for a range of masses. This is plausibly indicative of a connection between X-ray luminosity and assembly history at fixed mass. We will investigate the origin of scatter in future work. Once again it is remarkable that RG~572 does not stand out dramatically, its core-excised X-ray luminosity is only slightly lower than the rest of the sample. Since the use of Planck $M_{500}$ shifts the MCXC clusters away from the \cite{Arnaud:2010} relation, we expect that hydrostatic mass bias will play an important role when comparing the RG clusters to observations.

To conclude our comparison with X-ray scaling relations, in Figure~\ref{fig:MTX}, we show the mass--temperature relation for a subset of the MCXC clusters for which we could take the X-ray temperatures from the ACCEPT cluster catalog. Due to the known biases in estimating temperatures from simulated clusters that reflect temperatures measured from X-ray spectroscopy \citep[see e.g.][for in-depth comparisons]{Biffi:2014}, we show a range of differently weighted temperature estimates in the different panels of the figure. First, we compare with the spectroscopic-like temperature as introduced by \cite{Mazzotta:2004} as
\begin{equation}
T_{sl} = \frac{\int n^2 T^{\alpha-1/2}{\rm d}V}{\int n^2 T^{\alpha-3/2}{\rm d}V},\quad\textrm{with }\alpha=0.75,
\label{eq:tsl}
\end{equation}
as a fit to spectroscopic temperature estimates with {\em Chandra} or {\em XMM-Newton}. We additionally do not include cells with a temperature below 0.5~keV in this average, which is necessary in the case of cooling to avoid the inclusion of non-X-ray emitting gas. Furthermore, we again exclude the core inside $0.15\,R_{500}$ in order to avoid even larger fluctuations in the estimated cluster temperature due to the influence of the AGN model on this region. This is also common practice in observations to reduce scatter in the scaling relations \citep[e.g.][]{Mantz:2010}. We found that a temperature estimated by computing the mean energy of all photons emitted using an APEC model (see discussion above) above 0.1~keV leads to basically identical results within the scope of this first analysis, so that we do not consider these `mock' X-ray observations in this paper further. 

The other two temperature estimates we consider are a simple mass and a volume-weighted average of all cells {\em with no core excluded and no cut in minimum gas temperature}, so that they reflect the true mean temperatures, not including possible observational biases.

We note that the spectroscopic-like (as an APEC emissivity weighted temperature not shown here) temperature has a much larger variance  compared to the volume- and mass-weighted averages. The variance substantially increases if the core is also included due to the $n^2$-dependence of the emissivity. In all cases, and thus plausibly not affected by potential biases compared to observational results, the $M-T$ scaling that we observe is slightly steeper than the one exhibited by the MCXC/ACCEPT data. In particular, we find that the cluster temperature tends to be somewhat low for masses above $10^{14}\,M_\odot$. The discrepancy, while systematic, is not dramatic. It is consistent with  our comparison of the temperature profiles with the ACCEPT data, in which we found a similar temperature discrepancy at large radii. 

To summarise, in no case did we find a very strong dependence of any cluster scaling relation on the assembly history. A weak dependence might be possible for the $M-L_X$ relation in our data. This result is consistent with earlier analysis in this direction, by e.g. \cite{Jeltema:2008}, who however found a dependence on the dynamical state once the scaling relations are not compared against the true mass but against, e.g., the hydrostatic mass. This is very plausibly the case for our simulations as well, since the hydrostatic mass bias can be large close to $R_{500c}$ (see our analysis in Section~\ref{sec:hydrostaticmass}). It is thus not entirely clear if and how the ACCEPT comparison may be biased by hydrostatic mass errors which could alleviate also the discrepancies we observed between the profiles in Section~\ref{sec:gas_xray}. 


\section{Discussion of the numerical modelling approaches}
\label{sec:discussion}
We have established already in the previous sections that our results are numerically converged by comparing the ten runs at 4K resolution to the higher resolution 8K run of RG~653. Next, we investigate the dependence of our results on particular choices of the AGN feedback model parameters and discuss discrepancies with published results based on Lagrangian methods.


\subsection{AGN model dependence: X-ray properties}
\label{sec:agn_impact}
In this section, we investigate the robustness of our results to changes in the thermal AGN feedback model. In particular, we want to see whether the cold cores we find are stable to such changes. We furthermore will investigate the degree of change in the scaling relations that can result from such astrophysical changes beyond the resolution of the simulation in the subgrid model.

\begin{figure}
\begin{center}
\includegraphics[width=0.9\columnwidth]{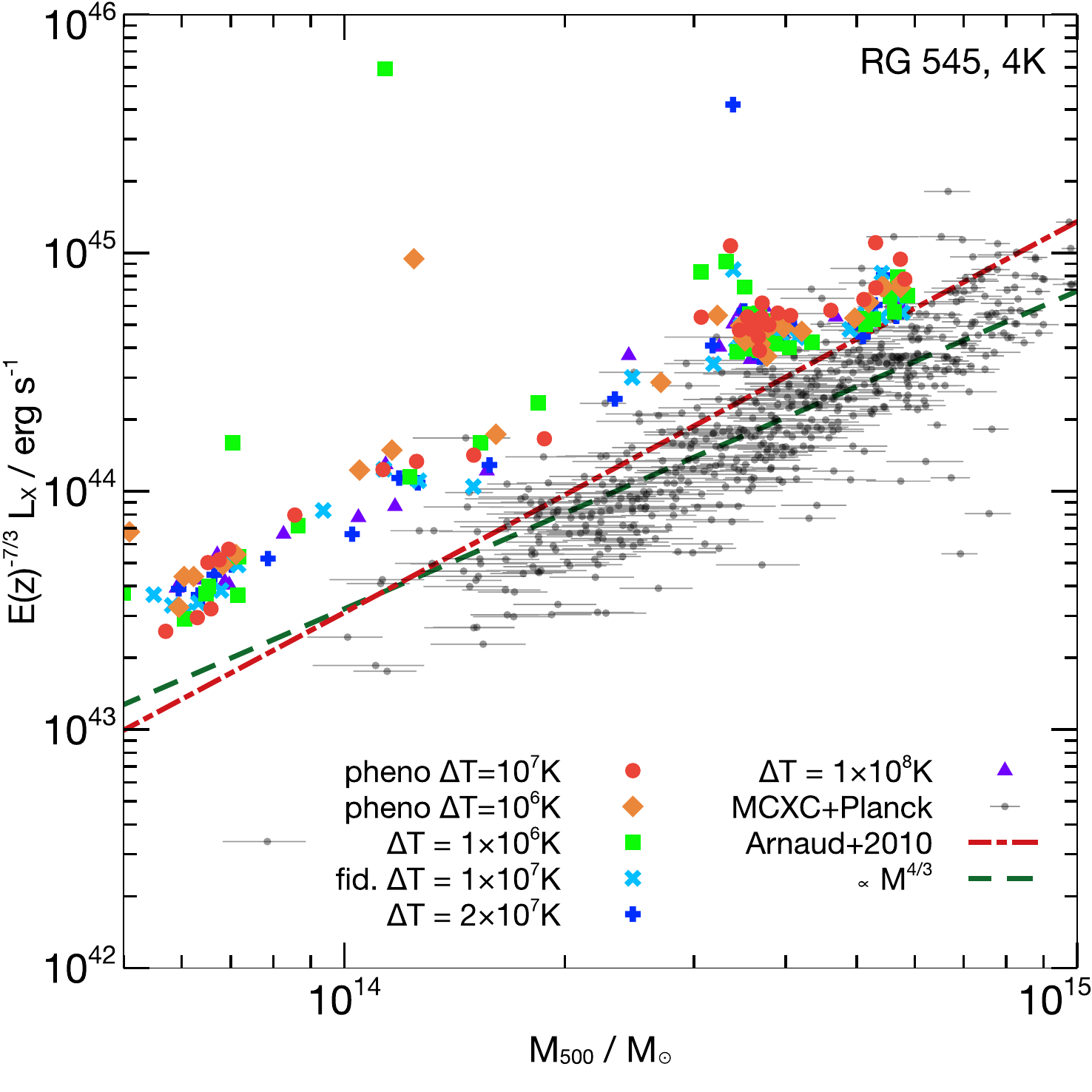}
\end{center}
\caption{\label{fig:MCXC_AGN}Impact of the AGN model on X-ray scaling relations: As Figure~\ref{fig:MCXC}, but only for the cluster RG545 using two different implementations and a range of parameters of the thermal blastwave AGN model. Specifically, we compare the phenomenological injection model ('pheno') that has a well-resolved injection bubble with the standard \citet{Booth:2009} model with injection close to the resolution scale. In both cases we vary the thermal energy accumulation threshold $\Delta T$. No significant impact is seen on the bolometric core-excised X-ray luminosity.}
\end{figure}

While it is clear that the very short cooling times in the centres of massive clusters necessitate a mechanism to prevent efficient cooling in their cores in order to reproduce both the observed gas fractions and realistic galaxy masses, what this mechanism is, or how it should be modelled on the scales accessible to three-dimensional cosmological simulations is less clear. Virtually all current state-of-the-art cosmological simulations that tap into this mass regime invoke some form of AGN feedback to resolve this problem. As shown in Figure~\ref{fig:gas_depletion}, the AGN model we adopted fails however to reduce the gas fraction at radii $\gtrsim 0.5\,R_{vir}$, where our simulations predict somewhat high baryon fractions compared to e.g. \cite{Mantz:2014}. If AGNs however have such a dramatic effect on the entire ICM, most of cluster cosmology would have to rely on a very tight relation between cluster mass and AGN feedback energy. The analysis of \cite{LeBrun:2014} e.g. shows that the entire baryon gas profile can be varied by tuning the energy accumulation threshold $\Delta T$ of the feedback model of \cite{Booth:2009}. While this is possible, it should be expected that AGN feedback could be a {\em significant} source of scatter, relating processes at the pc scale to the Mpc-scale of the ICM. In this section we thus repeat this analysis and investigate how sensitive our results are to variations of the energy accumulation parameter $\Delta T$.

In Figure~\ref{fig:MCXC_AGN}, we show again the evolution of the X-ray luminosity of a cluster as a function of its mass over time. For this analysis, we focus on the cluster RG~545, which we found to be a cool-core system during the time it is in the mass-range between $4<M_{500}/10^{14}M_{\odot}<6$. The selection of a CC cluster for this analysis will allow us to investigate later how robust the existence of the cool core in the simulation is when parameters of the AGN model are varied. We have not found any large differences between the CC and non-CC systems in neither the $M-L_X$ or $M-T$ scaling relations once there cores are excluded.

When changing $\Delta T$ over two orders of magnitude from $10^6$ K to $10^8$ K, we see no clear systematic effect on the X-ray luminosity.  This appears in tension with what \cite{LeBrun:2014} find (their Figure~1b), where the X-ray luminosity decreases by $~0.7-0.8$~dex when $\Delta T$ is increased by a similar amount. We caution that we only studied the response of a single system here, while \cite{LeBrun:2014} study the impact on the whole sample of clusters in the Cosmo-OWLS volume of 400 $h^{-1}$Mpc. As expected, the phenomenological injection model (labeled `pheno' in the figures, cf. Section~\ref{sec:agn_modelling}) is less bursty in X-ray luminosity compared to the fiducial injection method, since the energy is distributed over a larger volume. In conclusion, it does not appear possible in this model to tune the normalisation of the $L_X-M$ relation in any significant way using AGN feedback parameters in our simulation.

The excess X-ray luminosity is however consistent with the findings of e.g \cite{Choi:2015}, who find that in their simulations of group-scale haloes with thermal AGN feedback, the X-ray luminosity is a factor of $\sim50-100$ higher than observed. This excess luminosity is reduced, and becomes consistent once they employ a kinetic feedback model.


\subsection{AGN model dependence: stability of cool cores and gas depletion}
\label{sec:coldcore_agn}

\begin{figure}
\begin{center}
\includegraphics[width=0.9\columnwidth]{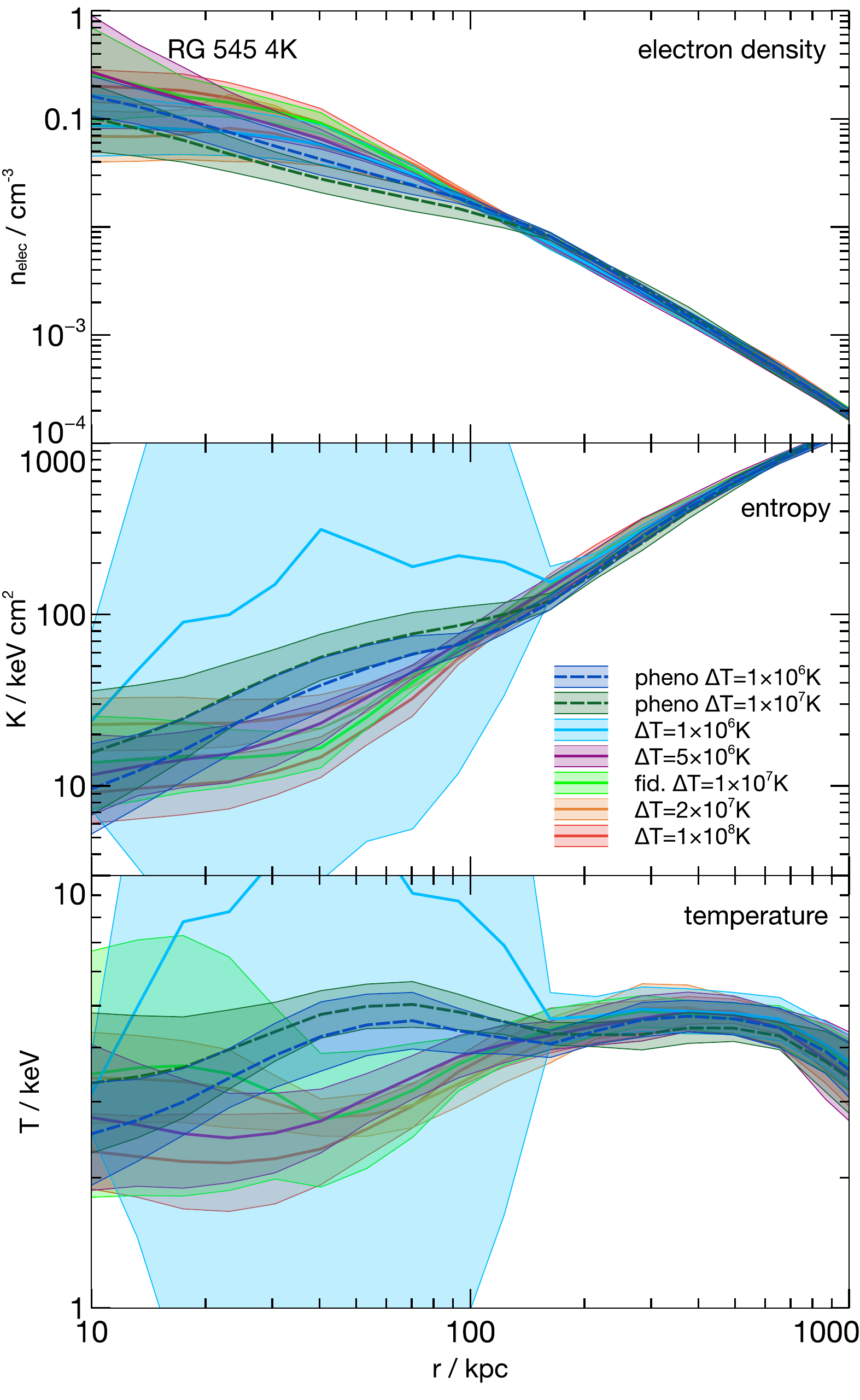}
\end{center}
\caption{\label{fig:profiles_AGN}Impact of AGN model on ICM profiles: As the top three panels of Figure~\ref{fig:accept_profiles}, but only for the CC cluster RG545 using two different implementations and a range of parameters of the thermal AGN injection threshold. The phenomenological model has the most significant impact by raising the entropy also at small radii, changing the electron density to a power law in the core, and flattening the temperature profile. The lowest injection threshold $\Delta T=10^6\,{\rm K}$ produces frequent outward travelling blast waves that show up in the entropy and temperature profile but only have a modest effect on the density. In no case radii outside  $\sim100\,{\rm kpc}$ are affected however. }
\end{figure}

Much like in our sample, \cite{Burns:2008} \citep[and][]{Planelles:2009} found the CC/NCC dichotomy to naturally arise in samples of cosmological cluster simulations. Unlike in RHAPSODY-G, their simulations were of much coarser resolution (AMR min cell size of $15.6\,h^{-1}{\rm kpc}$ and $N$-body particle mass of $9\times10^{9}\,h^{-1}{\rm M}_\odot$) and did not include AGN feedback which has left room for speculations about different origins of the CC/NCC dichotomy. Of course, this difference w.r.t. our simulations is crucial, since increasing the spatial resolution increases the severe central cooling catastrophe and leads to unrealistic BCG stellar masses providing the main argument for the necessity to include AGN feedback as a central energy source to compensate cooling losses and bring BCG masses into realistic ranges \citep[c.f.][]{Martizzi:2012,Martizzi:2012b}. It is thus a non-trivial result that the cool cores survive this energy injection, and one may wonder whether their survival is only due to particular choices of the feedback model parameters and whether more rare but violent, or more frequent but less violent AGN events might destroy them. This question is of particular importance not only because one may wonder about the model parameter dependence but also since the dominant role of AGNs in CC/NCC transitions has been advocated in the literature \citep{Guo:2009,Guo:2010}.

We next investigate the robustness of the cool core clusters to changes in the AGN energy injection threshold. We note once again that the injection threshold ${\Delta}T$ does not control the total energy injected into the ICM, but only its portioning. In Figure~\ref{fig:profiles_AGN}, we show the dependence of the electron density, temperature and entropy profiles on $\Delta{T}$ --- to be compared with Figure~\ref{fig:accept_profiles}. We find a non-monotonic dependence of the central core slope of the electron density profile on $\Delta T$. In particular, the lowest threshold we considered, $\Delta{T}=10^{6}$ K yields a cored profile with a central density of $\sim 0.1 {\rm cm}^{-1}$, more consistent with the ACCEPT observational constraints. At the same time, the frequent AGN bursts show up as outward travelling shock waves in the entropy and temperature profiles inside $\lesssim150\,{\rm kpc}$. In all cases considered there is no effect outside that radius on either entropy or density, nor is the entropy profile changed to that of the average NCC systems. This result lets us conclude that the formation of the cool core cannot be prevented by the central thermal AGN feedback model, regardless of the injection threshold and region we considered. The additional overcooling of the cool core can be somewhat tuned but not alleviated by this feedback model. Whether kinetic feedback can resolve this problem \citep[c.f.][]{Li:2015} is an interesting possibility to be investigated in future work.

\begin{figure}
\begin{center}
\includegraphics[width=0.9\columnwidth]{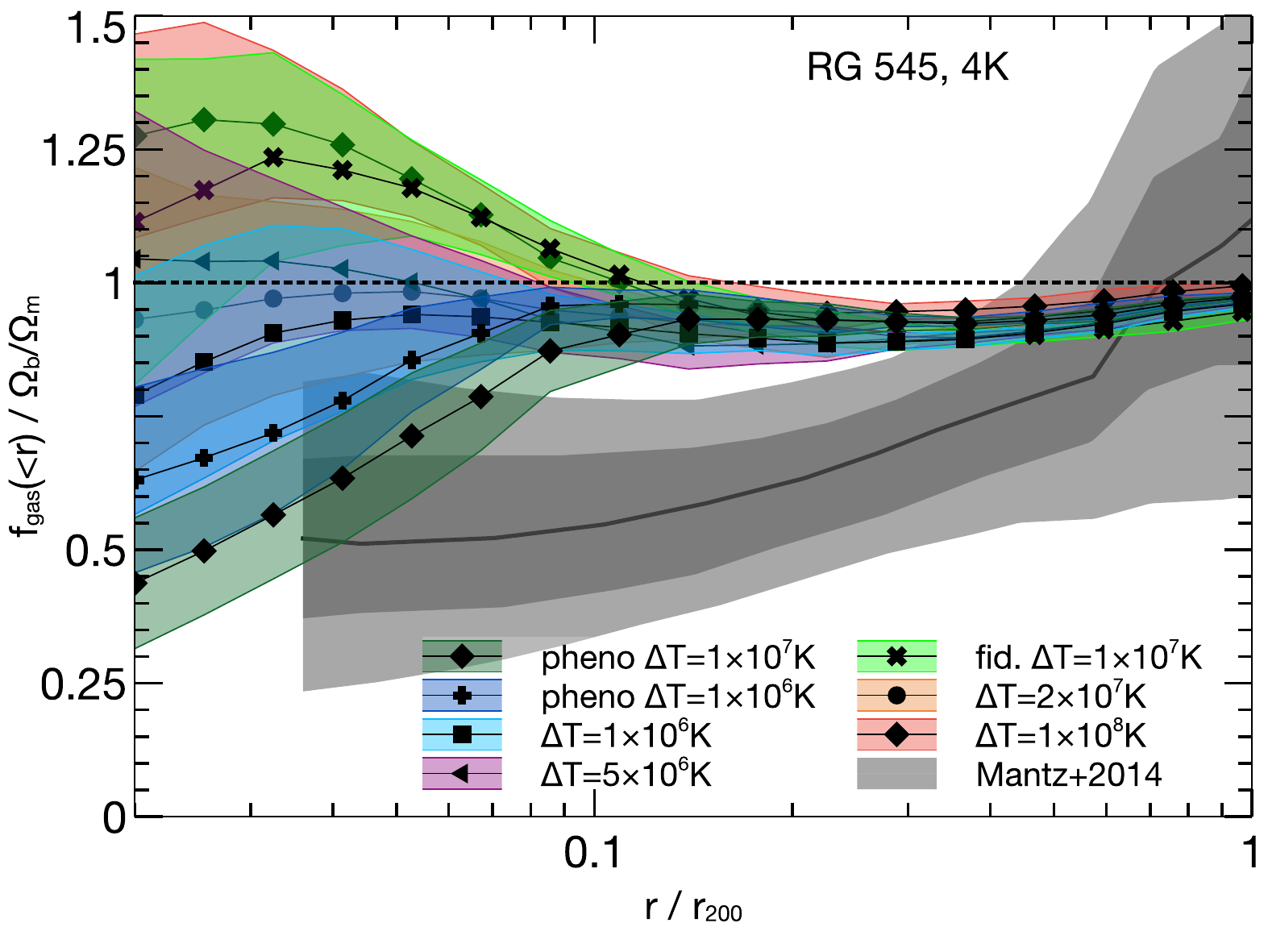}
\end{center}
\caption{\label{fig:fgas_AGN}Impact of AGN model on gas fraction: As Figure~\ref{fig:gas_depletion}, but only for the CC cluster RG~545 using two different implementations and a range of parameters of the thermal blastwave AGN model. The phenomenological model has the most significant impact to keep the gas fraction below the universal baryon fraction also at small radii, but in all cases the gas fraction at $\gtrsim 0.1R_{\rm vir}$ is high compared to the constraints from \citet{Mantz:2014}. In contrast to the non-cool-core case, the thermal blast wave here is insufficient to substantially reduce the gas fraction in the inner parts of the halo.}
\end{figure}

A robust property of our cool core systems that we identified above was the high central gas fraction of order the universal baryon fraction inside the cool cores --- inconsistent with the observations of e.g. \cite{Mantz:2014}. We thus also want to investigate the impact of the AGN model parameters on the gas depletion profiles. In Figure~\ref{fig:fgas_AGN}, we show essentially the same plot as in Figure~\ref{fig:gas_depletion}, varying again the injection threshold parameter $\Delta T$. While we saw that the lowest injection threshold $\Delta{T}=10^{6}$ K was able to impact the entropy and density profiles, it also reduces the central gas fraction, but only very slightly so, and only at radii $\lesssim R_{\rm vir}/10$. Similarly, the larger injection regions of the phenomenological model reduce the central gas fraction below the universal value. Once again, in no case could we affect regions outside the core leaving the gas fraction inconsistent with observations at $\sim0.2\,R_{\rm vir}$. The energy available from the thermal blastwave, even after accumulating energy to heat to higher temperatures is completely insufficient to reduce the gas fraction outside the core to the $f_{\rm gas} \sim 0.5 \Omega_b/\Omega_m$ observed by \cite{Mantz:2014}. Plausibly, thermal conduction might play a role here to distribute energy better towards the outskirts.


\subsection{A note on fundamental differences between AMR and SPH}
Various authors using SPH simulations have argued that the parameter $\Delta T$ has a strong impact on the physical properties of both the galaxies and the intracluster gas. As we have discussed above, this is in contrast to our own findings. \cite{LeBrun:2014} find changes in the gas density, \cite{Pike:2014} in the gas pressure out to $R_{500}$. Furthermore, \cite{LeBrun:2014} found that $\Delta T=10^8\,{\rm K}$ allowed them to make cluster properties compatible with a range of observables. We discussed in more detail the impact of the two flavours of energy injection as well as the energy accumulation threshold on the various observables and various physical properties in Section~\ref{sec:agn_impact} above. As we demonstrated there, in our Eulerian AMR simulations, the particular choice of $\Delta T$ and even the size of the injection region does not affect larger radii. It is thus plausible that thermal feedback has a different impact in Lagrangian and Eulerian simulations. Unfortunately, to the best of our knowledge, no direct comparison exists beyond the indirect estimate of \cite{Chaudhuri:2013} who find that the amount of energy per unit mass that the AGN has to provide is larger for SPH than for AMR. While the lack of entropy cores in purely adiabatic SPH simulations has been known for a long time \citep[see][for the original result and a recent explanation of its origin]{Frenk:1999,Power:2014}, more modern formulations of SPH are able to alleviate this discrepancy with Eulerian methods \citep[see e.g.][for recent method comparisons]{Rasia:2014,Sembolini:2015}. At the same time, the effect of feedback models in controlled experiments is not well documented. We hope that a comparison of the numerical discrepancies, as suggested by our results, will be undertaken by the community in the near future.

\subsection{A note on metal enrichment}
We find (see paper 2, Martizzi et al. 2015, in prep. for more details) that both our intracluster medium as well as the galaxy population have a metallicity which is slightly lower than what is of observed in the gas \citep[e.g.][]{DeGrandi:2004,Matsushita:2011,Werner:2013} and stellar metallicity \citep[e.g.][]{Gallazzi:2005}. A metallicity distribution more consistent with observations has been reproduced in various SPH simulations \citep[e.g.][]{Sijacki:2007,Planelles:2014}. Also in the {\sc Illustris} simulation (not SPH) realistic metallicities were produced \citep{Vogelsberger:2014}, albeit at the price of baryon fractions that are completely inconsistent with observational constraints \citep[][their Figure 1]{Haider:2015}. On the other hand, our results are consistent with the low metallicities found by other AMR simulations, e.g. \cite{Dubois:2014}. This aspect is another possible systematic discrepancy between the methods and should be investigated in more detail. We note that standard SPH does not include any mixing of metals (or any tracers, although explicit metal diffusion can be added, see e.g. \citealt{Shen:2010}). \cite{Planelles:2014} included a smoothed metallicity estimate when calculating cooling times and state that their unsmoothed metallicity field is very noisy. We investigate the aspect of metal enrichment in our simulations further in paper 2, Martizzi et~al. 2015, in prep. 


\section{Summary and Conclusions}
\label{sec:conclusions}
We present simulations of nine clusters of mass $M_{\rm vir}\sim10^{15}\,{\rm M}_\odot$ and one of twice that mass, including cooling, star formation, as well as supernova and AGN feedback with a physical resolution of $3.8\,h^{-1}{\rm kpc}$. The simulations include the environment of the clusters inside of spheres of $8\,h^{-1}{\rm Mpc}$ around each cluster at $z=0$. In this paper, we compare in detail the ICM density, temperature, entropy, and gas depletion profiles with X-ray data by performing a time-ensemble analysis for each cluster over a narrow mass bin and comparing with observed clusters in the same mass range. Next, we investigate the evolution of our simulated clusters over cosmic time with a range of cosmological observables that serve as mass proxies. In this paper, we focus on the mass vs Compton-$y$, X-ray luminosity and X-ray temperature scaling relations that are of particular importance for cluster cosmology. We also establish the numerical convergence of our results with resolution and their robustness against changes in the AGN feedback parameters.
\newline

\noindent We summarise our findings as follows:
\begin{enumerate}
\item We find a persistent cool-core/non-cool core dichotomy in our clusters. The cool cores are insensitive to changes in the thermal AGN feedback model parameters.

\item We link the disruption of cool cores to low-angular momentum major mergers. Major mergers with enough angular momentum leave the cool cores intact. Core disruption occurs on time scales of at most a few 100~Myr, with much increased core cooling times after the disruption, leading to a quick transition and thus a stable bimodality. 

\item Our simulations agree with the {\em Planck} $M_{500}-Y_{500}$ scaling relation with very little scatter. We do not identify a strong dependence of the scatter on the accretion history or the AGN model parameters.

\item The RG clusters are more X-ray luminous than a comparison sample from the MCXC catalog. The clusters evolve along scaling relations in the $M-L_{X}$ plane that are consistent with self-similar scaling. There is a slight indication that the scatter in this relation correlates with details of the assembly history.

\item The non-cool core clusters reproduce density, entropy and mass profiles of an ACCEPT comparison sample well, and are consistent with the observed gas depletion profiles. The cool core systems have excess central gas and a too low central entropy compared to the ACCEPT clusters. In addition, there is a general indication that at large radii, the simulated clusters have a slightly too low entropy and temperature, and a slightly too high density compared to observations.

\item The galaxies forming in our simulations have realistic masses and are consistent with abundance matching results across three decades in halo mass. At higher masses, the simulated galaxies are slightly more massive than observed at a given halo mass, although we caution that observational issues complicate a detailed comparison.  The star formation rates at the high mass end are consistent with recent observational constraints for BCGs. 
\end{enumerate}

\noindent The discrepancies we observe and listed above are plausibly related to shortcomings of the simplistic central thermal blast wave model:
\begin{enumerate}
\item In our AMR simulations, we find that thermal AGN feedback does not affect the ICM at significantly large radii.  We see no effect on gas at scales of $\sim R_{\rm vir}/2$, nor is the AGN able to mildly stabilise the cool core systems. Once a cool core forms, it cools below observed core entropies and leads to a core baryon fraction inconsistent with observational constraints.

\item The above finding is discrepant with SPH results in the literature. In our simulations, details of the AGN energy injection are irrelevant for global ICM properties. In particular, the X-ray and SZ scaling relations are unaffected by details of the thermal AGN model. This is quite in contrast to the findings of, e.g., \cite{LeBrun:2014} and points to a possible discrepancy between SPH/Lagrangian and Eulerian methods and how feedback couples to gas in such simulations.
\end{enumerate}

\noindent The inability of the thermal AGN model to shape larger scales in our simulations plausibly points to other forms of energy injection \cite[e.g. through kinetic feedback, see][]{Li:2015,Choi:2015} or additional processes in shaping the intracluster medium. Several published results suggest that thermal conduction might play a central role both in stabilising cool cores and at larger scales \citep[e.g.][]{Guo:2008,Parrish:2009,Ruszkowski:2010,Arth:2014}. We will investigate these aspects in future research.


\section*{Acknowledgements}

\noindent We thank Yohan Dubois, James Bartlett and Pawel Biernacki for discussions. We are grateful to Peter Behroozi for kindly making his {\sc Rockstar-Galaxies} code available to us for further modifications, and for providing data shown in Fig.~9 and 10. 

\noindent O.H. acknowledges support from the Swiss National Science Foundation (SNSF) through the Ambizione fellowship. HW acknowledges the support by the U.S. Department of Energy under contract number DE-FG02-95ER40899.  DM acknowledges support from the Swiss National Science Foundation. RHW received support from the U.S. Department of Energy contract to SLAC no. DE-AC02-76SF0051. This work was supported by a grant from the Swiss National Supercomputing Centre (CSCS) under project ID s416.



\appendix

\begin{table*}
\begin{center}
\begin{tabular}{l|c|c|c|c|c|c|c|c|c}
\hline
cluster name & $z$ & $M_{500}^{\rm MCXC}$ & $R_{500}^{\rm MCXC}$ & $M_{\rm SZ}^{\rm Planck}$ & $T_{cl}$ & $K_0$ & $L^{\rm MCXC}_{X,500}$ & $E^{-2/3}Y_{500}$ & cool core? \\
\hline
ABELL 85 & 0.0558 & 5.32 & 1.21 & 4.92 & 6.9 & 12.5 & 5.10 & 1.70 & + \\
ABELL 141 & 0.23 & 4.72 & 1.10 & 5.67 & 5.31 & 205.03 & 5.16 & 19.97 & - \\
ABELL 267 & 0.23 & 4.93 & 1.11 & 5.04 & 6.79 & 168.56 & 5.53 & 13.99 & - \\
ABELL 399 & 0.0716 & 4.25 & 1.12 & 5.24 & 5.8 & 153.2 & 3.59 & 4.08 & - \\
ABELL 586 & 0.171 & 5.20 & 1.16 & 5.17 & 8.7 & 94.75 & 5.62 & 6.23 & - \\
ABELL 611 & 0.288 & 4.60 & 1.06 & 5.50 & 6.69 & 124.93 & 5.33 & 64.94 & - \\
ABELL 907 & 0.1527 & 5.03 & 1.14 & 5.41 & 5.04 & 23.38 & 5.30 & 5.85 & - \\
ABELL 963 & 0.2056 & 4.73 & 1.11 & 5.83 & 6.6 & 55.77 & 5.03 & 13.02 & - \\
ABELL 1413 & 0.1426 & 5.55 & 1.19 & 5.95 & 8.9 & 64.03 & 6.04 & 6.71 & - \\
ABELL 1650 & 0.0843 & 4.12 & 1.10 & 4.45 & 5.89 & 37.96 & 3.47 & 1.75 & - \\
ABELL 1651 & 0.084 & 4.39 & 1.13 & 5.07 & 7.0 & 89.46 & 3.85 & 2.64 & - \\
ABELL 1664 & 0.1276 & 4.06 & 1.08 & 4.28 & 3.5 & 14.4 & 3.57 & 3.53 & + \\
ABELL 1795 & 0.0625 & 5.53 & 1.22 & 4.47 & 7.8 & 18.99 & 5.48 & 0.94 & + \\
ABELL 1995 & 0.3186 & 5.87 & 1.14 & 4.92 & 8.6 & 374.35 & 8.28 & 21.18 & - \\
ABELL 2034 & 0.113 & 4.07 & 1.09 & 5.85 & 7.15 & 232.64 & 3.51 & 3.17 & - \\
ABELL 2069$^\dagger$ & 0.116 & 4.57 & 1.13 & 5.31 & 7.9 & 453.25 & 4.26 & 8.29 & - \\
ABELL 2104 & 0.1554 & 4.42 & 1.10 & 5.74 & 9.31 & 160.61 & 4.23 & 15.39 & - \\
ABELL 2111 & 0.23 & 4.66 & 1.09 & 5.73 & 8.02 & 107.36 & 5.05 & 16.17 & - \\
ABELL 2244 & 0.0967 & 4.49 & 1.13 & 4.38 & 5.57 & 57.58 & 4.05 & 3.15 & - \\
ABELL 2294 & 0.178 & 4.23 & 1.08 & 5.98 & 7.1 & 156.31 & 4.05 & 6.87 & - \\
ABELL 2409 & 0.1479 & 4.63 & 1.12 & 5.06 & 5.5 & 73.81 & 4.53 & 6.75 & - \\
ABELL 3364 & 0.1483 & 4.50 & 1.11 & 4.89 & 6.59 & 268.55 & 4.32 & 4.99 & - \\
ABELL 3571 & 0.0391 & 4.51 & 1.15 & 4.63 & 7.6 & 79.31 & 3.82 & 0.63 & - \\
ABELL 3827 & 0.0984 & 4.59 & 1.14 & 5.77 & 8.05 & 164.58 & 4.20 & 3.79 & - \\
CL J1226.9+3332 & 0.89 & 4.39 & 0.83 & 5.70 & 10.4 & 166.03 & 11.25 & 181.25 & - \\
MS 0735.6+7421 & 0.216 & 4.60 & 1.09 & 5.02 & 5.45 & 15.96 & 4.87 & 10.28 & + \\
MS 0906.5+1110 & 0.163 & 4.74 & 1.12 & 5.42 & 8.1 & 104.23 & 4.86 & 7.21 & - \\
RXCJ0331.1-2100 & 0.188 & 4.05 & 1.06 & 4.34 & 4.61 & 11.4 & 3.82 & 5.36 & + \\
ZWCL 1358+6245 & 0.328 & 4.61 & 1.05 & 4.81 & 7.2 & 20.67 & 5.62 & 29.16 & + \\
\hline
${}^\dagger$ = excluded & & $\times10^{14}{\rm M}_\odot$ & Mpc & $\times10^{14}{\rm M}_\odot$ & {\rm keV} & ${\rm keV }{\rm cm}^2$& $10^{44}\,{\rm erg}/{\rm s}$ & ${\rm kpc}^2$\\
\hline
\end{tabular}
\end{center}
\caption{Catalog of ACCEPT clusters that have been selected for comparison with our simulated clusters. The clusters occupy the mass range between $4$ and $6\times10^{14}\,{\rm M}_\odot$ in $M_{500}$ based on both X-ray mass estimates as given in the MCXC catalog and SZ mass estimates as given in the {\em Planck} 2015 SZ union catalog. In this table, we list additional properties of these clusters as compiled from the ACCEPT, MCXC and {\em Planck} 2015 union catalogs. Compton-$y$ $Y_{5R500}$ have been converted from the {\em Planck} SZ catalog quantities given in arcmin$^2$ to kpc$^2$  assuming $h = 0.7, \Omega_m = 0.3$ and $\Omega_\Lambda = 0.7$ in the angular diameter distance relation.}
\label{tab:accept_sample}
\end{table*}%

\section{ACCEPT cluster comparison subsample}
In Table~\ref{tab:accept_sample}, we list the key properties of the subset of the ACCEPT clusters that we use for comparison.

\label{lastpage} \end{document}